\documentclass[11pt, reqno,preprint]{article}
\pdfoutput=1
\usepackage{jheppub}
\usepackage{epsfig}
\usepackage{amssymb}
\usepackage{amsmath}
\usepackage{mathrsfs}
\usepackage{hyperref}
\usepackage{multirow}

\def\be{\begin{equation}}
\def\ee{\end{equation}}
\def\ba{\begin{eqnarray}}
\def\ea{\end{eqnarray}}
\def\nl{\nonumber\\}
\def\aa{a}

\def\Res{\mbox{Res}}
\def\ZZ{\mathcal{Z}}

\def\Eq#1{Eq.~(\ref{#1})}
\def\Li{\textrm{Li}}
\def\l{\langle}
\def\r{\rangle}

\title{Jumpstarting the all-loop S-matrix of planar $\mathcal{N}=4$ super Yang-Mills}
\author[a]{S. Caron-Huot,}
\author[b]{Song He}
\affiliation[a]{School of Natural Sciences, Institute for Advanced
Study, Princeton, NJ 08540, USA} \affiliation[b]{Max-Planck-Institut
f\"ur Gravitationsphysik, Am M\"uhlenberg 1, 14476 Potsdam, Germany}
\emailAdd{schuot@ias.edu, songhe@aei.mpg.de} \abstract{We derive a
set of first-order differential equations obeyed by the S-matrix of
planar maximally supersymmetric Yang-Mills theory. The equations,
based on the Yangian symmetry of the theory, involve only finite and
regulator-independent quantities and uniquely determine the all-loop
S-matrix. When expanded in powers of the coupling they give
derivatives of amplitudes as single integrals over lower-loop,
higher-point amplitudes/Wilson loops. We outline a derivation for
the equations using the Operator Product Expansion for Wilson loops.
We apply them on a few examples at two- and three-loops, reproducing
a recent result on the two-loop NMHV hexagon and fixing previously
undermined coefficients in a recent Ansatz for the three-loop MHV
hexagon. In addition, we consider amplitudes restricted to a
two-dimensional subspace of Minkowski space, and obtain some equations which involve only that sector.}
\begin{document}
\maketitle

\pagebreak

\section{Introduction}

$\mathcal{N}=4$ supersymmetric Yang-Mills theory (SYM) is believed
to be integrable in the planar limit~\cite{minahanzarembo,
beisertstaudacher}, cf.~\cite{intreview} for a recent review. This
has made it possible to compute quantities of the theory, such as
the spectrum of anomalous dimensions, at any value of the
coupling~\cite{BDS0, BES, Ysystem}. On the other hand, remarkable
structures in the S-matrix of the theory have been unraveled recently. 
Among them, a hidden, dual superconformal symmetry has been discovered both at strong~\cite{aldaymalda} and weak coupling~\cite{magic} for the
S-matrix, which, together with the ordinary superconformal symmetry, generates an infinite-dimensional symmetry encoding the integrability of
the theory, the so-called Yangian symmetry\footnote{This was proposed originally in ~\cite{Yangian}, and the form of the Yangian algebra used in
this paper, which takes dual superconformal algebra as the level-zero subalgebra, has been discussed in ~\cite{Drummond:2010qh}.}.

The dual superconformal symmetry can be understood as the symmetry of null polygonal Wilson loops in a dual spacetime. The bosonic Wilson loops
are dual to maximally-helicity-violated (MHV) scattering amplitudes at strong~\cite{aldaymalda} and weak
coupling~\cite{Drummond:2007aua,WL1,WLDrummond,WLDrummond2}, and recently the duality has been generalized to the case of arbitrary helicity
(N${}^k$MHV) amplitudes and supersymmetric Wilson loops~\cite{masonskinner,arXiv:1010.1167} (or with a closely related light-cone limit of
correlation functions \cite{reg1,reg2}). Although generally tree amplitudes are Yangian
invariant~\cite{sokatchevDCI,brandhubernote,Grassmannian,Grassmannian2}, the naive Yangian symmetry is broken for loop-level amplitudes/Wilson
loops even if we consider finite quantities, such as the remainder and ratio functions~\cite{sokatchevDCI, sokatchevsDCI, arXiv:1105.5606}. In
this paper, we will argue that the Yangian symmetry can be made exact for all-loop amplitudes/Wilson loops, which are in turn completely
determined by the all-loop equations derived from the exact symmetry.

In order to discuss regulator-independent relations in a uniform
way, it proves convenient to introduce the {\it BDS-subtracted}
S-matrix element $R_{n,k}$, \be A_{n,k} = A_n^\textrm{BDS} \times
R_{n,k}\label{defR} \ee where $A_{n,k}$ stands for the N${}^k$MHV
scattering amplitude and $A_n^\textrm{BDS}$ for the exponentiated
Ansatz proposed by Bern, Dixon and Smirnov (BDS)~\cite{Bern:2005iz},
including the MHV tree and coupling constant factor.

The BDS-subtracted S-matrix $R_{n,k}$ is infrared finite and
regulator independent. It is invariant under the action of a chiral
half of the dual superconformal symmetry as well as under dual
conformal transformations. In this paper we propose a compact,
all-loop equation for the action of other dual superconformal
generators, denoted as $\bar Q$, in terms of a one-dimensional
integral over the collinear limit of a higher-point amplitude: \be
\bar Q^A_a R_{n,k}= \aa ~\Res_{\epsilon=0}
\int_{\tau=0}^{\tau=\infty}
\left(d^{2|3}\mathcal{Z}_{n{+}1}\right)_a^A \left[ R_{n{+}1,k{+}1} -
R_{n,k} R^\textrm{tree}_{n{+}1,1}\right]+\rm{cyclic}, \label{Qbar}
\ee where $a,A=1,\ldots,4$ are momentum-twistor indices and
$\epsilon,\tau$ parametrize $\mathcal{Z}_{n{+}1}$ in the collinear
limit ($\tau$ being related to the longitudinal momentum fraction),
Eq.~(\ref{expansion}). In this equation, $\aa=\aa(g^2)$ is one
quarter of the cusp anomalous dimension \be
 \aa:=\frac14 \Gamma_\textrm{cusp} = g^2 -\frac{\pi^2}{3} g^4 + \frac{11\pi^4}{45}g^6+\ldots,
\ee known exactly at all values of the coupling $g^2=\frac{g^2_\textrm{YM}N_c}{16\pi^2}$~\cite{grisha,grisha2,BES}. We expect the equation to be
exact at any value of the coupling, but in this paper we will study it perturbatively with respect to $\aa$.

We find \Eq{Qbar} natural and pleasing in many respects.
First, it relates finite and regulator-independent quantities.
Integrating out a particle with measure
$(d^{2|3}\mathcal{Z})^A_a$ is virtually the simplest operation one
could imagine, which carries the quantum numbers of $\bar Q$.
The one-dimensional collinear integral
over $\tau$ reflects the physical intuition that naive $\bar Q$ is violated because it causes
asymptotic states to radiate collinearly.  The presence of two
terms on the right-hand side has a simple explanation: if the
first term is viewed as the effect of $\bar Q$ on the amplitude,
then the term with $R^\textrm{tree}_{n{+}1,1}$ is due to the
action on the $A^\textrm{BDS}$ factor in \Eq{defR}.
The proportionality to $\Gamma_\textrm{cusp}$ of the second term is thus easy to understand,
it being the constant of proportionality in the BDS Ansatz,
while the structure itself is rigid: certain divergences which would violate conformal invariance
cancel between the two terms.
The fact that only $1\to 2$ splitting appears to all loop
orders, as opposed to $1\to 3,4,\ldots$ seems difficult to understand from the scattering amplitude viewpoint,
and we can only derive it through the duality with Wilson loops.

The equation holds for generic configurations, that is, it neglects
so-called distributional terms which are supported on singular
configurations. These terms were used in~\cite{arXiv:0905.3738} to
determine tree amplitudes. By stripping off the MHV tree, which give
rise to such terms, the tree amplitudes have been argued to be
uniquely determined by requiring analytic properties such as the
right collinear behavior, in addition to Yangian
invariance~\cite{sokatchevsDCI}. In this spirit, we will assume that
all the pertinent information is included by imposing in addition to
\Eq{Qbar} the correct collinear limits of BDS-subtracted amplitudes,
which play the role of boundary conditions to \Eq{Qbar}.

\begin{figure}\centering
\includegraphics[height=2cm]{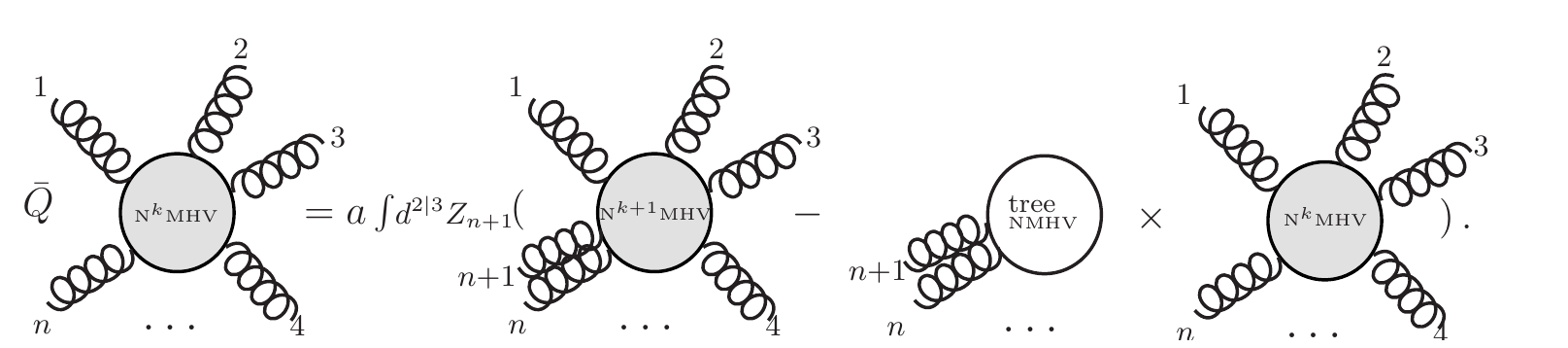}
\caption{All-loop equation for planar $\mathcal{N}=4$ S-matrix.} \label{fig:equation}
\end{figure}

Using the discrete parity symmetry of scattering amplitudes, we can derive an equivalent equation
for the level-one generator \be
 Q^{(1)a}_A R_{n,k}= \aa~Z_n^a \lim_{\epsilon\to 0} \int_0^\infty \frac{d\tau(d\eta_{n{+}1})_A}{\tau}
 \left(R_{n{+}1,k} - \sum_{1\leq j<i\leq n{-}3}C_{n,i,j}\frac{\partial R_{n,k}}{\partial \chi_j}\right) + \textrm{cyclic},
\label{Q1}
\ee
where $C_{n,i,j}$ is given in \Eq{Cnij}.
The level zero generators $\bar Q$ and $Q$ together with $Q^{(1)}$ generate the full Yangian algebra.

It is significant that the right-hand sides of \Eq{Qbar} and \Eq{Q1}
take the form of linear operators acting on the BDS-subtracted
S-matrix (viewing $R_{n,1}^\textrm{tree}$ as a collection of
constants, see \eqref{NMHVtree}). This means that the right-hand
side could be moved to the left, and the resulting operators
interpreted as quantum-corrected $\bar Q$ and $Q^{(1)}$ which
annihilate the S-matrix. In other words, these equations are not
``anomaly equations'' --- their content is precisely that all
Yangian anomalies can be removed by a simple redefinition of the
generators. Symmetry generators which receive quantum corrections
but nonetheless admit simple closed form expressions are not
uncommon in integrable systems, see for instance~\cite{luscher}.
This being said, we will continue to write these equations in the
form of a naive (or bare) generator on the left, with the correction
on the right-hand side, as this will prove most useful for our
applications.

The form of (\ref{Qbar}) is very similar to that obtained by Sever
and Vieira in the context of a proposed CSW-regularization of
amplitudes~\cite{anomaly2}. The essential new features are the focus
on the finite quantity $R$, which gives rise to the differenced
form, and the advantage of working with integrated amplitude, which
results in an all-loop relation with an overall proportionality to
$\Gamma_\textrm{cusp}$. Our formulas also reproduce the one-loop
results of~\cite{Beisert:2010gn}.

Our derivation of \Eq{Qbar} will be based on the Operator Product Expansion
(OPE) for null polygonal Wilson
loops~\cite{Alday:2010ku}.  It will be supported by an explicit computation of the
fermion dispersion relation to $\mathcal{O}(\Gamma_{\textrm{cusp}}^2)$, finding agreement with ~\cite{basso}. We will also
present strong explicit evidence for its all-loop validity, through two- and
three-loop computations. In particular, we have reproduced the
two-loop MHV~\cite{Goncharov:2010jf} and NMHV
hexagon~\cite{Dixon:2011nj}, and obtained new results for NMHV
heptagon and three-loop MHV hexagon, where we have fixed the two
undetermined coefficients in the recent Ansatz for its
symbol~\cite{Dixon:2011pw}.

The paper is organized as follows. We start in
section~\ref{sec:review} with a short review on momentum twistors,
BDS Ansatz and Yangian symmetry. In section~\ref{sec:Qbar}, we
explain how to use~\eqref{Qbar} to compute $\bar Q$, especially at
one loop, how are MHV and NMHV amplitudes uniquely determined
by~\eqref{Qbar}, and how to derive~\eqref{Q1} from it. We then
employ the method to reproduce results for two-loop MHV amplitudes
in section~\ref{sec:twoloopmhv}. We continue in
section~\ref{sec:NMHV} to derive results for two-loop NMHV hexagon
and heptagon, as well as three-loop MHV hexagon. In
section~\ref{sec:anomalousdim}, we outline a derivation of the
equation. We also discuss two-dimensional kinematics in
section~\ref{sec:2d}. We finish with some conclusions and appendices
containing techniques and details of some computations.

\section{Momentum twistors, BDS Ansatz and Yangian
symmetry}\label{sec:review}

Since our discussion will center on the dual superconformal
symmetry, it is advantageous to use momentum-twistor variables
introduced by Hodges~\cite{hodges}, which manifest the symmetry at
least for tree amplitudes. The Wilson loop dual to $n$-point
amplitude is formulated along a $n$-sided null polygon in a chiral
superspace with coordinates $(x,\theta)$, \be x_i^{\alpha
\dot\alpha}-x_{i{-}1}^{\alpha
\dot\alpha}=\lambda_i^{\alpha}\bar\lambda_i^{\dot\alpha},\qquad
\theta_i^{\alpha A}-\theta_{i{-}1}^{\alpha
A}=\lambda_i^{\alpha}\eta_i^A, \ee for $i=1,\ldots,n$, where
$\alpha, \dot\alpha$ are SU(2) indices of spinors $\lambda_i$ and
their conjugates $\bar\lambda_i$ encoding the null momenta of $n$
particles,  and $A$ is the SU(4) index of Grassmann variables
$\eta_i$ describing their helicity states. The momentum (super-)
twistors are defined as \be \mathcal{Z}_i=
(Z_i^a,\chi_i^A):=(\lambda_i^{\alpha},x_i^{\alpha \dot\alpha}
\lambda_{i \alpha},\theta_i^{\alpha A} \lambda_{i \alpha}).\ee We
further define the totally antisymmetric contraction of four
twistors, $\l i j k l\r:=\varepsilon_{abcd}Z^a_iZ^b_jZ^c_kZ^d_l$,
and the basic R-invariant of five super-twistors, \be
[i\,j\,k\,l\,m]:=\frac{\delta^{0|4}(\l\l i\,j\,k\,l\,m\r\r)}{\l i j
k l\r \l j k l m\r\l k l m i\r\l l m i j \r\l m i j k\r},\ee where
the argument of Grassmann delta function is $\l\l
i\,j\,k\,l\,m\r\r^A:=\chi_i^A\l j k l m\r+\textrm{cyclic}$. NMHV
tree (divided by MHV tree), appearing in~\eqref{Qbar}, is simply
given by a sum of these R-invariants \be\label{NMHVtree}
R^{\textrm{tree}}_{n,1}=\sum_{1<i<j<n}[1\,i\,i{+}1\,j\,j{+}1].\ee

At loop level, the symmetry of amplitudes is broken by infrared divergences, which need to be regulated and subtracted for exact symmetry. Based
on the known infrared behavior and the ABDK iterative relation~\cite{Anastasiou:2003kj}, BDS have proposed an exponentiated Ansatz for all-loop
MHV amplitudes in $D=4-2\epsilon$ dimensions~\cite{Bern:2005iz}, \be
\frac{A_n^{\textrm{BDS}}}{A^{\textrm{tree}}_{n,\textrm{MHV}}}=1+\sum^{\infty}_{\ell=1}\tilde{g}^{2\ell}
M^{(\ell)}_n(\epsilon)=\exp\left[\sum^{\infty}_{\ell=1}\tilde{g}^{2\ell}\left(f^{(\ell)}(\epsilon)M^{(1)}_n(\ell\epsilon)+C^{(\ell)}+E^{(\ell)}_n(\epsilon)\right)\right],\ee
where $\tilde{g}^2:=2 g^2(4\pi e^{-\gamma})^{\epsilon}$ has been used as the parameter of loop expansion, $f^{(\ell)}(\epsilon)=\frac14
\Gamma^{(\ell)}_\textrm{cusp}+\mathcal{O}(\epsilon)$, $C^{(\ell)}$ are independent of kinematics or $n$ (non-vanishing for $\ell>1$), and
$E_n^{(\ell)}$ vanish as $\epsilon\to 0$; by stripping off the MHV tree $A^{\textrm{tree}}_{n,\textrm{MHV}}=\frac{\delta^4(\sum_i \lambda_i\bar
\lambda_i)\delta^{0|8}(\sum_i\lambda_i\eta_i)}{\l 1 \, 2\r\l 2\, 3\r\ldots\l n\, 1\r}$,  the one-loop amplitude
$M^\textrm{1-loop}_n:=A^\textrm{1-loop}_{n,\textrm{MHV}}/A^{\textrm{tree}}_{n,\textrm{MHV}}$ is given by \be M^\textrm{1-loop}_n=-\frac
1{2\epsilon^2}\sum^n_{i=1} \left(-\frac{x^2_{i,i{+}2}}{\mu^2}\right)^{-\epsilon}+F^\textrm{1-loop}_n(\epsilon),\ee where
$F^\textrm{1-loop}_n(\epsilon)$ is a sum of finite parts of the so-called two-mass easy box functions~\cite{Bern:2005iz}.
The BDS Ansatz is believed to be exact for $n=4,5$, in which case
$R_{4,0}=R_{5,0}=R_{5,1}/R_{5,1}^\textrm{tree}=1$, and $R$ behaves
simply under collinear limits, both $k$-preserving $R_{n,k}\to
R_{n{-}1,k}$ and $k$-decreasing $\frac{\int d^4\chi_n R_{n,k}}{\int
d^4\chi_n [n{-}2\,n{-}1\,n\,1\,2]}\to R_{n{-}1,k{-}1}$, as
$\mathcal{Z}_{n}\to \mathcal{Z}_{n{-}1}$ (using a parametrization such as \Eq{expansion}).


As a consequence of Poincar\'e supersymmetry of scattering amplitudes, the BDS subtracted amplitude is invariant under a chiral half of dual
superconformal symmetry and the R-symmetry, and it is also believed to be invariant under all bosonic symmetries, including dual conformal
symmetry, \be Q^a_A=(Q^\alpha_A, \bar S^{\dot\alpha}_A):=\sum^n_{i=1} Z_i^a\frac{\partial}{\partial \chi_i^A},\quad
R^A_B:=\sum^n_{i=1}\chi^A_i\frac{\partial}{\partial \chi_i^B},\quad K^a_b:= \sum^n_{i=1} Z_i^a\frac {\partial }{\partial Z_i^b}.\ee Naively the
BDS-subtracted amplitude is not annihilated by generators in the other chiral half, \be\label{Qbardef} \bar Q^A_a=(S^A_\alpha, \bar
Q^A_{\dot\alpha}):=\sum^n_{i=1}\chi_i^A \frac{\partial}{\partial Z_i^a}, \ee but as we can see from~\eqref{Qbar}, the symmetry is restored by a
quantum-corrected $\bar Q$. Note the correction is manifestly $Q$-invariant, thus the $Q$ invariance and \eqref{Qbar} imply the invariance under
$K^a_b=\frac12\{Q^a_A,\bar Q^A_b\}$. For Yangian symmetry, one needs at least one additional level-one generator, e.g. $Q^{(1) a}_A$ which
contains the ordinary superconformal generator $s^{\alpha}_A:=\sum^n_{i=1}\frac{\partial}{\partial \lambda_{i \alpha}} \frac{\partial}{\partial
\eta_i^A}$, \be
  Q^{(1)a}_A =\frac12 \left(\sum_{i<j}-\sum_{j<i}\right) \left( Z_i^a \frac{\partial}{\partial Z_i^b} Z_j^b \frac{\partial}{\partial \chi_j^A}
   -Z_i^a \frac{\partial}{\partial \chi_i^B} \chi_j^B \frac{\partial}{\partial \chi_j^A}\right).
\ee
Note that, although second order in derivatives, this is only first order in bosonic derivatives.
Since $s^{\alpha}_A$ is the parity conjugate of $\bar
s_{\dot\alpha}^A$, which is part of $\bar Q$, we will derive the
Eq.(\ref{Q1}) from the conjugate of Eq.(\ref{Qbar}).

\section{The $\bar Q$ equation}\label{sec:Qbar}

In this section we elaborate on the evaluation of \Eq{Qbar}. It involves adding a particle in a collinear limit. In the case of edge $n$,
we parameterize its (super-)twistor as
\be
 \mathcal{Z}_{n{+}1} = \mathcal{Z}_n - \epsilon \mathcal{Z}_{n{-}1} + C\epsilon\tau \mathcal{Z}_1
+ C'\epsilon^2 \mathcal{Z}_2, \label{expansion} \ee with $C=\frac{\l
n{-}1n 23\r}{\l n123\r}$ and $C'=\frac{\l n{-}2\,n{-}1\,n1\r}{\l
n{-}2\,n{-}1\,21\r}$. The collinear limit is $\epsilon\to 0$ and,
physically, $\tau$ is related to the momentum fraction shared by
particle $n{+}1$ in that limit.  The most general collinear limit would require three parameters, and the third one, which we will not need,
could be obtained by replacing $C'$ with an order one function.
The signs and normalization have been chosen such that, for real $\epsilon>0$ and
$\tau>0$, Euclidean $n$-gons (configurations with
positive cross-ratios) are approached by Euclidean $(n{+}1)$-gons.

The basic operation $\mbox{res}_{\epsilon=0}\int
(d^{2|3}\ZZ_{n{+}1})^A_a$ can be evaluated as follows. In our
parametrization, the bosonic part of the measure is
$(d^2Z_{n{+}1})_a:=(Z_{n{+}1}dZ_{n{+}1}dZ_{n{+}1})_a=C(n{-}1n1)_a
\epsilon d\epsilon d\tau$, where only the dominant part at
$\epsilon\to 0$ was kept. Thus \be
 \mbox{res}_{\epsilon=0}\int_{\tau=0}^{\tau=\infty} (d^{2|3}\ZZ_{n{+}1})^A_a  = C (n{-}1n1)_a  ~
  \mbox{res}_{\epsilon=0} \int \epsilon d\epsilon \int_0^\infty d\tau (d^3\chi_{n{+}1})^A.
\ee The notation $\mbox{res}_{\epsilon=0}$ means to extract the
coefficient of $\frac{d\epsilon}{\epsilon}$ in the $\epsilon\to 0$
limit. It is not trivial that this exists, but we will see that
there are never singularities stronger than
$\frac{d\epsilon}{\epsilon}\log^{\ell-1}\epsilon$ in $\bar Q$ of the
$\ell$-loop amplitude, and that the logarithms always go away after
$\tau$ integration.

\subsection{R-invariants}

It is useful to illustrate the procedure on the simplest non-trivial
object, NMHV R-invariants. If the R-invariant does not involve
$\ZZ_{n{+}1}$, the Grassmann integral will produce zero.
Furthermore, even if $\ZZ_{n{+}1}$ appears, a pole $1/\epsilon$ will
be absent unless $\mathcal{Z}_n$ is also present. Thus the only
R-invariants which give non-trivial results contain both $\ZZ_n$ and
$\ZZ_{n{+}1}$.

Consider the invariant $[i\,j\,k\,n\,n{+}1]$ for $i,j,k$ all
distinct from $n{-}1$ and $1$. After doing the $\chi_{n{+}1}$
integration one gets \be
 \int (d^{2|3}\ZZ_{n{+}1})^A_a [i\,j\,k\,n\,n{+}1]
 = C(n{-}1n1)_a \int \frac{\epsilon d\epsilon d\tau\l\l i\,j\,k\,n\,n{+}1\r\r^A\l ijkn\r^2}{\l ijnn{+}1\r\l jk nn{+}1\r\l ki nn{+}1\r\l
 ijkn{+}1\r},
\ee and plugging in the parametrization \Eq{expansion} and keeping
the dominant term as $\epsilon\to 0$ gives \ba
\mbox{res}_{\epsilon=0}\int_{\tau=0}^{\tau=\infty}
(d^{2|3}\ZZ_{n{+}1})^A_a [i\,j\,k\,n\,n{+}1] &=&
 C (n{-}1n1)_a \mbox{res}_{\epsilon=0} \frac{d\epsilon}{\epsilon}\int_0^\infty d\tau\frac{\l\l i\,j\,k\,n\,B\r\r^A\l ijkn\r}{\l ij\,nB\r\l jk nB\r\l ki nB\r}
 \nl &=& C(n{-}1n1)_a \int_0^\infty d\tau \frac{\l\l i\,j\,k\,n\,B\r\r^A\l ijkn\r}{\l ijnB\r\l jk nB\r\l ki nB\r}, \label{Rinv1}
\ea where $\ZZ_B:=\ZZ_{n{-}1}-C \tau \ZZ_1$.  We could perform the
$\tau$ integral here, but it is advantageous not to do so and keep
the $\tau$-integrand untouched at this stage. This is because in
later applications we will need this integral with additional
dependence on $\tau$ inserted. However, we can simplify it a bit. It
has three poles hence two linearly independent residues. Define the
bitwistor $X=X(\tau):=n\wedge B$. Then the residue at $\l ij X\r=0$
gives \be
 (n{-}1n1)_a \frac{\l\l i\,j\,k\,n\,[n{-}1\r\r^A\l1]ijn\r\l ijkn\r}{\l ijn1\r\l jkn[n{-}1\r\l1]ijn\r\l kin[n{-}1\r\l1]ijn\r}
  = (n{-}1n1)_a \frac{\l\l n{-}1\,n\,1\,i\,j\r\r^A}{\l n{-}1n1i\r\l n{-}1n1j\r}.
\ee This can be rewritten using the nice identity \be (n{-}1n1)_a
\frac{\l\l n{-}1\,n\,1\,i\,j\r\r^A}{\l n{-}1n1i\r\l n{-}1n1j\r}
 \nl = \bar Q^A_a \log \frac{\l \bar ni\r}{\l \bar nj\r},
\ee where $(\bar n):=(n{-}1 n 1)$. Similarly the residue at $\l ik X\r=0$ is $\bar Q \log \frac{\l \bar n i\r}{\l\bar n k\r}$, and the residue
at $\l jk X\r=0$ is given by minus the sum of the two. Adding these contributions we have,\be
 \mbox{res}_{\epsilon=0}\int_{\tau=0}^{\tau=\infty} d^{2|3}\ZZ_{n{+}1}[i\,j\,k\,n\,n{+}1]
 = \int_0^\infty\left(
   d \log \frac{\l Xij\r}{\l Xjk\r} \bar Q\log \frac{\l \bar n j\r}{\l \bar n i\r}
 + d \log \frac{\l Xjk\r}{\l Xik\r} \bar Q\log \frac{\l \bar n k\r}{\l \bar n
 i\r}\right).
\ee This is valid at the level of the $\tau$-integrand, for
$i,j,k\neq n{-}1,1$. Other R-invariants are computed similarly, and
we complete this subsection by giving the result of the
$\mbox{res}_{\epsilon=0}\int_{\tau=0}^{\tau=\infty}
d^{2|3}\ZZ_{n{+}1}$ operation on these R-invariants, with  $i,j\neq
n{-}1,1$, \ba \!\![i\,j\,n{-}1\,n\,n{+}1]
 &\to&
  \int d\log \frac{\l Xij\r}{\l Xn{-}2n{-}1 \r} \bar Q\log \frac{\l \bar n j\r}{\l \bar n i\r},
 \nl
~\!\![i\,j\,n\,n{+}1\,1]
  &\to&
  \int d\log \frac{\l Xij\r}{\l X12 \r} \bar Q\log \frac{\l \bar n j\r}{\l \bar n i\r},
  \nl
~\!\![i\,n{-}1\,n\,n{+}1\,1]
  &\to& \int  d\log \frac{\l Xn{-}2n{-}1\r}{\l X12 \r} ~\bar Q \log \frac{\l \bar n 2\r}{\l \bar n i\r}.
 \label{Rinv}
\ea
All other R-invariants giving zero. Note these expression all hold
at the level of the $\tau$-integrand.

\subsection{$\bar Q$ of one-loop amplitudes}

Armed with just this result, we are ready to evaluate the $\bar Q$
of any one-loop amplitude. The right-hand side of \Eq{Qbar} reads
\be
 \bar Q R^\textrm{1-loop}_{n,k} = \mbox{res}_{\epsilon=0}\int_{\tau=0}^{\tau=\infty} d^{2|3}\ZZ_{n{+}1}\left(
  R_{n{+}1,k{+}1}^\textrm{tree} - R_{n,k}^\textrm{tree}R_{n{+}1,1}^\textrm{tree}\right)+\textrm{cyclic}.
\ee Using the (P)BCFW formula for removing $\ZZ_{n{+}1}$ (associated
to the shift $\ZZ_{n{+}1}\to \ZZ_{n{+}1}+z
\ZZ_1$)~\cite{ArkaniHamed:2010kv},
the parenthesis can be rewritten as \be \sum_{i=2}^{n{-}2}
[n\,n{+}1\,1\,i\,i{+}1] \left(\sum_{k'=0}^k
R^\textrm{tree}_{i{+}2,k'}(\widehat{n{+}1},1,\ldots,i,\hat{I}_i)
R^\textrm{tree}_{n{+}1{-}i,k{-}k'}(\hat{I}_i,i{+}1,\ldots,n)
 - R_{n,k}^\textrm{tree}\right),
 \label{BCFW}
\ee up to $\ZZ_{n{+}1}$-independent terms which do not contribute to
the integral. The dependence on $\ZZ_{n{+}1}$ is in the R-invariant
and in the shifted twistors, $\widehat{n{+}1}:=(n n{+}1)\cap (1 i
i{+}1)$, $\hat{I}_i:=(i i{+}1)\cap (n n{+}1 1)$. However, the
parenthesis has a smooth collinear limit since no term depends
simultaneously on both $\ZZ_n$ and $\ZZ_{n{+}1}$. Thus we can merely
replace $\hat \ZZ_{n{+}1}$ by $\ZZ_n$ and $\hat{I}_i$ by its limit
$I_i=(n{-}1n1)\cap (ii{+}1)$ (supersymmetrically, the fermions of
$I_i$ are taken from $\chi_i$ and $\chi_{i{+}1}$). The $\ZZ_{n{+}1}$
dependence is limited to the R-invariant and using \Eq{Rinv} the
integral gives \be
 \bar Q R^\textrm{1-loop}_{n,k} = \int_{\tau=0}^{\tau=\infty} \sum^{n{-}2}_{i=2} d\log \frac{\l X ii{+}1\r}{\l X12\r} \bar Q \log \frac{\l \bar n i\r}{\l \bar n i{+}1\r}
 (\mbox{parenthesis in \Eq{BCFW}})+\textrm{cyclic}.\label{1loopanomaly1}
\ee In the parenthesis, $\widehat{n{+}1}\to n, \hat{I}_i\to I_i$, and nothing depends on $\tau$.

We must show that this integral is convergent at its endpoints. Near
$\tau=0$, there is a pole due to $d\log \l X n{-}2n{-}1\r$ in the
term $i=n{-}2$.  However, the two terms in the parenthesis cancel in
this case, so there is no problem.  There is also a pole at
$\tau=\infty$, due to $d\log \l X12\r$ present in every term. It is
nontrivial to see that it cancels out in the sum, but can be proved
as follows.

Instead of using the (P)BCFW formula associated with shifting
$\ZZ_{n{+}1}$, we could have used the BCFW formula associated with the
shift $\ZZ_n\to\ZZ_n+z \ZZ_{n{-}1}$. Then we would have obtained the same formula but with the
R-invariants replaced with $[i\,i{+}1\,n{-}1\,n\,n{+}1]$, and so
$d\log \frac{\l X ii{+}1\r}{\l Xn{-}2n{-}1\r}$ in the integrand, but
the parenthesis in the $\epsilon\to 0$ limit unchanged. This form
would make convergence at $\tau=\infty$ manifest but not at $\tau=0$. We
conclude that overall convergence follows beautifully from
the equality of the BCFW and (P)BCFW representations of tree
amplitudes.

Given these cancelations, we can integrate \Eq{1loopanomaly1} termwise by dropping the $d\log \l X12\r$ factor and the $i=n{-}2$ term, obtaining
simply \ba
 \bar Q R^\textrm{1-loop}_{n,k} &=& \sum_{i=2}^{n{-}3} \log \frac{\l n1ii{+}1\r}{\l n{-}1nii{+}1\r} \bar Q \log \frac{\l \bar n i\r}{\l \bar n i{+}1\r}
\nl &&\times
 \left(\sum_{k'}
R^\textrm{tree}(n,1,\ldots,i,I_i)R^\textrm{tree}(I_i,i{+}1,\ldots,n)-R_{n,k}^\textrm{tree}\right) + \textrm{cyclic},
\label{1loopanomaly} \ea which agrees with the formula
of~\cite{Beisert:2010gn} (there the product of tree amplitudes is interpreted in terms of
unitarity cuts). The subtraction of $R_{n,k}^\textrm{tree}$ arises
because we are considering the BDS-subtracted amplitude, which is
the one-loop N${}^k$MHV ratio function in this case.

\subsection{Uniqueness of $\bar Q$ solutions at MHV and NMHV}\label{sec:uniqueness}

The $\bar Q$ equation is especially interesting because as we will see now,
it fixes uniquely MHV \emph{and} NMHV amplitudes (assuming that the
right-hand side is known). This is not too difficult to see
for MHV amplitudes using the momentum-twistor form of $\bar Q$,
\eqref{Qbardef}. Indeed, taking derivatives of the equation $\bar Q f(Z)=0$
for any function of bosonic $Z$'s, $f(Z)$, we have,\be
 \frac{\partial}{\partial \chi_i^1} \bar Q^1_a f(Z) = 0 \Rightarrow \frac{\partial}{\partial Z_i^a} f(Z)=0.
\ee This equation, for all particle labels $i$ and twistor indices
$a=1\ldots 4$, implies that a bosonic function annihilated by $\bar
Q$ is a constant. Thus the ambiguity of the $\bar Q$ equation is at
most a constant, which can be fixed using the properties of the
BDS-subtracted amplitudes in collinear limits.

For NMHV amplitudes, we have to work harder to restrict the kernel of $\bar Q$.
A simple example which illustrates this at $5$-points is $ [12345] \log \frac{\l 1234\r}{\l 1235\r}$.
This has vanishing $\bar Q$ because
\be
 [12345]\bar Q\log \frac{\l 1234\r}{\l 1235\r}= [12345]\frac{(123)\l\l12345\r\r}{\l 1234\r\l1235\r}  \label{repofzero}
\ee contains $\l\l 12345\r\r$ both explicitly and from the Grassmann delta
function $\delta^{0|4}(\l\l 12345\r\r)$ in the R-invariant, hence
vanishes. On the other hand, this expression is not acceptable
because the argument of the logarithm is not conformal invariant
(synonymous with little group invariance in what follows): it has
non-vanishing weight with respect to 4 and 5. So it does not
correspond to any real ambiguity. This turns out to be general: any
NMHV expression with neutral little group and annihilated by both
$Q$, $\bar Q$, is a sum of R-invariants with constant coefficients.

To prove this, we first note that by $Q$ invariance alone, any NMHV
expression can be written as
\be
 F= \sum_{2\leq j<k<l<m\leq n} [1\,j\,k\,l\,m] F_{j,k,l,m}(Z)
\ee where the $\left(\begin{array}{c} n{-}1\\4\end{array}\right)$
$[1\,j\,k\,l\,m]$'s form a basis for all independent NMHV
R-invariants at $n$-point~\cite{freedman}. Each $F_{j,k,l,m}(Z)$ is
a conformal invariant function of the bosonic $Z$'s.

To show that the $F_{j,k,l,m}(Z)$ must be constant, we pick
$i\,\,/\!\!\!\!\!\!\!\in\!\! \{1,j,k,l,m\}$ and extract a specific component,
$\chi^1_i\chi^1_j\chi_k^2\chi_l^3\chi_m^4$, of $Z_j^a \bar Q^1_a F$.
The only way $\chi^1_j$ can arise is either from $\bar Q
F_{j,k,l,m}$ or from a R-invariant, but since $Z_j^a
\frac{\partial}{\partial Z_j^a} F=0$, it can only arise from a
R-invariant. Since only $[1\,j\,k\,l\,m]$ contains the prescribed
components, we deduce that \be Z_j^a \frac{\partial}{\partial
Z_i^a}F_{j,k,l,m}=0. \ee Repeating this with permutations of
$j,k,l,m$ shows that $F_{j,k,l,m}$ is independent of $Z_i$, 
and repeating for other $i$'s shows that $F_{j,k,l,m}$ depends only
on twistors $1,j,k,l,m$. But since there are no nontrivial
little-group invariant functions of five twistors, $F_{j,k,l,m}$
must be a constant, QED.

All remaining constant ambiguities can be fixed by collinear limits.
As mentioned, there are both $k$-preserving and $k$-decreasing
collinear limits. It turns out that just four of the $k$-preserving
limits suffice for any $n$.  For instance, working in the same
basis, the $k$-preserving collinear limit $\ZZ_1\to \ZZ_2$ will fix
all constants except those multiplying invariants of the form
$[1\,2\ldots]$. Taking the limit $\ZZ_2\to \ZZ_3$ will then fix the
coefficient of all but those beginning with $[1\,2\,3\ldots]$, and so
on.

These results open up the possibility of using the $\bar Q$ equation
to compute nontrivial MHV and NMHV amplitudes. Given one-loop
N${}^k$MHV amplitudes as the seed for recursion, this will restrict
the applications in this paper to two-loop MHV and NMHV and
three-loop MHV. To go beyond NMHV it becomes necessary to use both
the $\bar Q$ and $Q^{(1)}$ equations\footnote{A simple
counter-example to $\bar Q$ uniqueness is the invariant
$\frac{\delta^{0|4}(\l 1234\r \chi_5\chi_6 +
\textrm{cyclic})}{\l1234\r\cdots \l6123\r}$ which arises in the
6-point N${}^2$MHV tree amplitude and depends on six twistors. Any
conformal invariant cross-ratio of the six twistors multiplying it
will be $\bar Q$-invariant.}, or, equivalently, the $\bar Q$
equation and parity. Uniqueness then follows from a theorem proved
in \cite{arXiv:1002.4625,arXiv:1002.4622}: \emph{all Yangian
invariants are combinations of compact contour integrals inside the
Grassmannian $G(k,n)$}. We conclude that any N${}^k$MHV expression
annihilated by (naive) $Q,\bar Q, Q^{(1)}$ can only be a
combinations of such invariants, multiplied by c-numbers, which we
expect to be determined by collinear limits.

\subsection{The one-loop NMHV hexagon}
\label{sec:6pt1loop}

In the case $n=6$, equation (\ref{1loopanomaly}) evaluates more or less directly to
\be \bar Q
R_{6,1}^\textrm{1-loop} =
  ((5)+(3)) \log u_3\bar Q\log \frac{\l5612\r}{\l5613\r}
+ (1) \log u_3\bar Q\log \frac{\l 5613\r}{\l5614\r} +\mbox{cyclic},
\label{anomaly6a} \ee where $(1)$ is the R-invariant [23456], $(i)$
is obtained by a cyclic shift, and we have used that
$R_{6,1}^\textrm{tree}=(1)+(3)+(5)=(2)+(4)+(6)$.
The appearance of $\log u_3$ is easy to understand from the two terms in \Eq{1loopanomaly},
because they correspond to two poles of the $\tau$-integral, and so what multiplies the log has to be equal and opposite.
We use the following cross-ratios \be
 u_1=\frac{\l 1234\r\l 4561\r}{\l 1245\r\l 3461\r},\qquad
 u_2=\frac{\l 2345\r\l 5612\r}{\l 235
 6\r\l 1245\r},\qquad
 u_3=\frac{\l 3456\r\l 6123\r}{\l 3461\r\l 2356\r}.  \label{cr6}
\ee

We have just shown that the information in \Eq{anomaly6a} should suffice to determine
$R_{6,1}$.  Let us see how this works. The crucial step is to
bring the right-hand side to a form where the argument of each $\bar
Q$ is the logarithm of a conformal invariant cross-ratio; this form
will be unique. This can be achieved by adding suitable combinations
of zero in the form of equation (\ref{repofzero}), for which there
is a systematic procedure.

The following procedure reduces this to a simple linear algebra
problem. The first step is to remove the ambiguities in writing the
R-invariants, by using the identity $(1)-(2)+(3)-(4)+(5)-(6)=0$ to
remove $(6)$. We can then use four distinct nontrivial
representations of zero, $[(1)-(2)+(3)-(4)+(5)]$ times $\bar Q
\frac{\l 1234\r}{\l 1235\r}$, $\bar Q \frac{\l 1234\r}{\l 1245\r}$,
$\bar Q \frac{\l 1234\r}{\l 1345\r}$, $\bar Q \frac{\l 1234\r}{\l
2345\r}$, to remove e.g. the little group weight with respect to $i$
of the coefficient of $(i)$, for $i=1,\ldots 4$.

Actually, there is a final constraint: it is not trivial the little
group weight with respect to $5$ of the coefficient of $(5)$ is also
removed; but this is is the case. Then the coefficient of $(i)$ has
correct little group weight with respect to $i$ for $i=1,\ldots,5$.
The little-group weights with respect to other variables can then be
removed using equation (\ref{repofzero}) with R-invariants
$(1),\ldots, (5)$.

This procedure is simple to follow but not particularly
illuminating, so we spare the reader the details, recording only the
final result: \be \bar Q R_{6,1}^\textrm{1-loop} = \left(
R_{6,1}^\textrm{tree} \bar Q \log \frac{u_1u_2}{1-u_3}
 - ((1)+(4))\bar Q\log u_2 - ((2)+(5))\bar Q\log u_1\right)\log u_3 +\textrm{cyclic}.
\ee This equation is equal to \Eq{anomaly6a}, but now the $\bar Q$
acts on conformal invariants. The upshot is that in this form we are
allowed to directly integrate $\bar Q$: \be R_{6,1}^\textrm{1-loop}
=  R_{6,1}^\textrm{tree} (\log u_2\log u_3 +\Li_2(1-u_3)) -
((1)+(4)) \log u_2\log u_3  + \textrm{cyclic}+C, \label{NMHV6} \ee
where $C$ is an undetermined combination of R-invariants with
$c$-number coefficients.

To fix $C$, we can consider collinear limits. For instance, the
ratio function should vanish in the $k$-preserving limit $\ZZ_6\to
\ZZ_5$, corresponding to $u_1\to 0$ and $u_2\to 1-u_3$.  This limit
probes the coefficient of $(5)$ plus the coefficient of $(6)$. In
this limit, what we have in \Eq{NMHV6} goes to
$\frac{\pi^2}{3}[12345]$. All other $k$-preserving limits go to the
same number, allowing us to fix \be
 C=-\frac{\pi^2}{3}R_{6,1}^\textrm{tree}. \label{NMHV6C}
\ee
This is the correct ratio function!

\subsection{Derivation of equation (\ref{Q1}) from equation (\ref{Qbar})
}\label{sec:Q1}

This subsection lies a bit outside the main scope of this paper.
As noted in the Introduction, the equation for $\bar Q$ together with
parity symmetry of scattering amplitudes implies an equation for $Q^{(1)}$.

To derive it, the first step is to express \Eq{Qbar} in the language
of scattering amplitudes. We only need to do this for the two
components of $\bar Q$ which coincide with the ordinary
superconformal generators $\bar s^A_{\dot\alpha}$~\cite{Yangian}.
Technically, we really only need to do this for the first term in
the parenthesis of \Eq{Qbar}, and we can drop the explicit
dependence on $\epsilon$, reinstating it at the end. We find, after
reinstating the MHV prefactor and changing variable $C \tau\to
\frac{\l n1\r}{\l n{-}1n \r} x/(1-x)$, \be \bar s^A_{\dot\alpha}
A_{n,k}= \tilde \lambda_{n\dot\alpha} \lim_{\epsilon\to 0} \int_0^1
dx (d^3\chi)^A
   A_{n{+}1,k{+}1}(\ldots,\{\lambda_n,x\tilde \lambda_n,x\tilde \eta_n + \chi\},\{\lambda_n,(1{-}x)\tilde \lambda_n,(1{-}x)\tilde \eta_n - \chi\})
+\ldots\, .\nonumber \ee The variable $x$ is the usual longitudinal
momentum fraction. With the help of the {\it BCFW} computer package
for the evaluation of tree amplitudes \cite{arXiv:1011.2447}, we
have verified that the integral gives the correct result acting on
the NMHV 5,6,7 point tree amplitudes. The upshot is that in this
form it is possible to immediately write down the parity-conjugate
equation:\footnote{The easiest way to derive this equation is to
consider the case where particle $n$ is a positive-helicity gluon in
the $\bar s^A_{\dot\alpha}$ equation. Then the $\chi$ integral gives
eight terms on the right hand side, involving a minus-helicity
fermion, or a scalar plus a plus-helicity fermion, with various
R-symmetry assignments. Parity dictates that when $n$ is a
negative-helicity gluon, $s_A^\alpha$ should produce the eight
parity conjugate terms. The correctness of the other cases follows
by supersymmetry.}
\be
 s_A^\alpha A_{n,k} = \lambda_n^\alpha \lim_{\epsilon\to 0} \int_0^1 \frac{dx (d\chi)_A}{x(1-x)} A_{n{+}1,k}(\ldots,\{\lambda_n,x\tilde \lambda_n,x \tilde \eta_n + \chi\},
 \{\lambda_n,(1{-}x)\tilde \lambda_n,(1{-}x)\tilde \eta_n -
 \chi\})+\ldots\,,
\nonumber \ee  where the denominator $1/x(1-x)$ comes from a little
group transformation needed after interchanging $\lambda$ and
$\tilde \lambda$. The final step is to convert this equation back to
momentum twistors: \be
 Q^{(1)a}_A A_{n,k}= Z_n^a \lim_{\epsilon\to 0}\int_0^\infty \frac{d\tau}{\tau} (d\chi_{n{+}1})_A R_{n{+}1,k}(1,\ldots,n{+}1) + \ldots
\ee where we have put back the $\epsilon$ dependence, $\ZZ_{n{+}1}(\epsilon,\tau)$
being again given by \Eq{expansion}. Strictly speaking, $s^\alpha_A$
gives two out of the four twistor components of $Q^{(1)a}_A$.  The
remaining two components come for free, because the level-zero conformal
symmetry of Wilson loops is unbroken acting on BDS-subtracted amplitudes.

This takes care of the first term in the parenthesis of \Eq{Qbar}.
To deal with the second term we need the explicit form of acting
$\bar Q$ on NMHV tree,

\be \mbox{res}_{\epsilon=0} d^{2|3}\ZZ_{n{+}1}
R^\textrm{tree}_{n{+}1,1}=\sum_{i=2}^{n-3} \frac{\l\l \bar
n\,i\,i{+}1\r\r}{\l \bar n\, i\r\l \bar n \,i{+}1\r} d\log \frac{\l
n\,n{+}1\,i\,i{+}1\r}{\l n\,n{+}1\,n{-}2\,n{-}1\r}.\ee

We take its parity conjugate using $\chi_i=\sum_{j=n{-}2}^i \l
ij\r\eta_j$ and then $\eta_j\to \frac{\partial}{\partial \eta_j}$.
In terms of momentum twistors, the end result is \be
 Q^{(1)a}_A R_{n,k}= \aa~Z_n^a \lim_{\epsilon\to 0} \int_0^\infty \frac{d\tau}{\tau}
 \left((d\chi_{n{+}1})_A R_{n{+}1,k} - \sum_{1\leq j<i\leq n{-}3}C_{n,i,j}\frac{\partial R_{n,k}}{\partial \chi^A_j}\right) + \textrm{cyclic} \label{Q11}
\ee
where
\be
 C_{n,i,j}(\tau) = \frac{\l n \bar i\r\l j\overline{i{+}1}\r- \l n \overline{i{+}1}\r\l j\bar i\r}{\l n \bar i\r\l n \overline{i{+}1}\r} \tau \frac{d}{d\tau} \log \frac{\l nn{+}1 ii{+}1\r}{\l nn{+}1 n{-}2n{-}1\r},
\label{Cnij}
\ee
which is the formula recorded in the Introduction.

Because this is a consequence of unbroken parity symmetry and the equation for $\bar Q$, this does not require separate verification.
To ascertain that \Eq{Q11} contains no mistake, we have tested it on known expressions for 1-loop 6,7,8-point NMHV amplitudes.

\section{Two-loop MHV amplitudes}\label{sec:twoloopmhv}

Armed with just the $\bar Q$ equation (\ref{Qbar}), expressions for
one-loop NMHV ratio functions, and the $d^{2|3}\ZZ$ integral of
R-invariants \Eq{Rinv}, we are now ready to analyze two-loop MHV
amplitudes.

\subsection{The square and the pentagon}
\label{sec:pentagon}

In the cases $n=4$ and $n=5$, it is well-known that $\log
R_{4,5}=0$~\cite{Bern:2005iz}. Let us begin by reproducing this
simple result starting from equation (\ref{Qbar}). For $n=4$, this
is essentially trivial for all loops provided $\log R_5=0$ at one
lower loop order, so the equation reads \be
 \bar Q R_4=0.
\ee This implies that $R_4$ is a constant, which must be trivial by the
boundary condition.

In the case $n=5$, starting at two loops, the right-hand side is not
so trivially zero. Rather, it involves the collinear limit of the
one-loop six-point NMHV amplitude given in~\Eq{NMHV6}. Letting \be
 \mbox{res}_{\epsilon=0}\int_{\tau=0}^{\tau=\infty} d^{2|3}\ZZ_6 R_6^\textrm{1-loop}= \bar Q\log \frac{\l 4512\r }{\l 4513\r} \times I,
 \ee we
get that the R-invariants contribute to $I$ as follows, \be
 (1)\to d\log \frac{\tau}{\tau+1}, \quad (2)\to d\log \tau,\quad (4)\to -d\log (\tau+1),\quad (3),(5),(6)\to 0. \label{Rinv5}
\ee In the collinear limit $u_1\to \epsilon^2$, $u_2\to
\frac{1}{1+\tau}$, $u_3\to \frac{\tau}{1+\tau}$, allowing us to
write $I=I_1+I_2$ with \ba
 I_1 &=& \int_0^\infty d\left(\log \frac{(1+\tau)^2}{\tau}\right) \log (1+\tau) \log(1+1/\tau), \nl
 I_2 &=& \log\epsilon^2 \times \int_0^\infty d\left(\log (1+\tau) \log(1+1/\tau)\right).  \label{5ptint}
\ea
We find that $I_1=I_2=0$, confirming that $\bar Q R_5=0$ as expected.
This is the first nontrivial hint that the equation is working beyond one-loop.

The reader might worry about the divergent prefactor
$\log\epsilon^2$ in front of $I_2$.  Shouldn't the $\epsilon\to 0$
limit entering our basic equation be well-defined? The answer is
that the order of operations is important. The limit $\epsilon\to 0$
will always be well-defined provided the integration over $\tau$ is
carried out first. If one were to take instead $\epsilon\to 0$ with
fixed $\tau$,  one would find a divergence. This divergence has a
simple explanation and is actually predicted by the Wilson loop
OPE~\cite{Alday:2010ku}. We will return to it in subsection
\ref{sec:nologepsilon} where we confirm the quantitative prediction
for it, and also give the general argument for its cancelation after
$\tau$-integration.

\subsection{The hexagon}

For $n=6$, we need the one-loop seven-point NMHV amplitude, which
can be put in a compact form~\cite{ArkaniHamed:2010gh}, \be
R^\textrm{1-loop}_{7,1}=
[1,\{2,3\},\{4,5,6\}]I_1+[1,\{2,3\},\{4,5,6,7\}]I'_1+\rm{cyclic},\ee
where \be
[i,\{i{+}1,...,j\},\{k,...,l\}]=\sum^{\{j-1,j\}}_{J=\{i+1,i+2\}}\sum^{\{l,k\}}_{K=\{k,k+1\}}[i,J,K],
\ee
\begin{align}
I_1=&\phantom{+}\Li_2(1-v_7v_3)+\Li_2(1-v_1)+\Li_2(1-v_3v_6)+\Li_2(1-v_6v_2)\nl
&-\Li_2(1-v_1v_4)-\Li_2(1-v_6)-\Li_2(1-v_3)-\Li_2(1-v_5v_1)+\log
v_7\log v_2,\nl
I'_1=&\phantom{+}\Li_2(1-v_7)+\Li_2(1-v_6)+\Li_2(1-v_3)+\Li_2(1-v_5v_1)\nl
&-\Li_2(1)-\Li_2(1-v_7v_3)-\Li_2(1-v_3v_6)+\log v_7\log v_6,
\label{7ptII'}\end{align} and $I_i, I'_i$ are obtained by cyclic
shifts for $i=2,...,7$. Here we need to define cross-ratios beyond
six points\be u_{i,j,k,l}=\frac{\l i\,i{+}1\,j\,j{+}1\r\l
k\,k{+}1\,l\,l{+}1\r}{\l i\,i{+}1\,k\,k{+}1\r\l
j\,j{+}1\,l\,l{+}1\r}, \ee and at seven points, a basis of
cross-ratios can be chosen as $v_i:=u_{i{+}1,i{+}3,i{+}4,i}$ for
$i=1,...,7$. In the collinear limit $\ZZ_7 \to \ZZ_6$,
$v_4\rightarrow 0$, $v_3\rightarrow (1-v_2)/(1-v_2 v_6)$ and
$v_5\rightarrow (1-v_6)/(1-v_2v_6)$, thus the result depends on
$v_1,v_2,v_6,v_7$.

The R-invariants appearing are not independent, and it is convenient
to choose those containing the label 2 as a basis. Upon doing the integral
over $d^{2|3}\ZZ_7$, only $[12367]$, $[12467]$, $[23467]$, $[23567]$
and $[24567]$ contribute ($[12567]$ does not contribute, because its
coefficient has to vanish due to the k-decreasing collinear limit constraint), which
produce, proceeding as in the five-point example, \be
 \bar Q R^\textrm{2-loop}_{6,0}=
   (I_{1,1}+I_{1,2}) \bar Q\log \frac{\l 5613\r}{\l 5612\r}
  +(I_{2,1}+I_{2,2})\bar Q\log \frac{\l 5614\r}{\l 5612\r}+\textrm{cyclic},  \label{Qbar6}
\ee
where
\ba
 \hspace{-1cm}
 I_{1,1} &=& \int_0^\infty\left(\begin{array}{l}
 \phantom{+} d\log(\frac{\tau}{\tau+u_3}) \left(  \log \frac{u_2 (\tau+u_3)}{\tau} \log \frac{u_3(\tau+1)}{\tau+u_3}+\Li_2(1-u_3)-\Li_2(\frac{1-u_3}{\tau+u_3})\right)
 \\
 + d\log(\tau+1) \left( \log u_1(\tau+1)\log \frac{\tau+u_3}{\tau+1} + \Li_2(1-u_3) -\Li_2(\frac{(1-u_3)\tau}{\tau+u_3})\right)
 \\
 + d\log(\frac{\tau+u_3}{\tau+1}) \left(\log \frac{u_2(\tau+u_3)}{u_3}\log \frac{\tau+1}{\tau+u_3}
  -\log (\tau+1)\log \frac{u_1(\tau+1)}{\tau+u_3} \right.
   \\ \hspace{2.5cm} \left.
  +\Li_2(1-u_1)+\Li_2(1-u_2)+\log u_1 \log u_2-\frac{\pi^2}{6}\right)
  \end{array}\right),\nl
     I_{1,2} &=&
 \log \epsilon^2 \times \int_0^\infty d\left( \log \frac{u_3(\tau+1)}{\tau+u_3}\log(\frac{\tau}{\tau+u_3}) + \log(\tau+1)\log \frac{\tau+u_3}{\tau+1}\right),
 \ea
and
\ba
 I_{2,1} &=& \int_0^\infty\left(\begin{array}{l}
\phantom{+}
  d\log \frac{\tau+u_3}{\tau} \left( \log\frac{u_2(\tau+u_3)}{\tau}\log \frac{u_3(\tau+1)}{\tau+u_3} -\log (\tau+1)\log \frac{\tau+1}{\tau}
  \right.\\ \hspace{2.5cm} \left.
  +\Li_2(1-u_2)+\Li_2(1-u_3)
    -\Li_2(1-\frac{u_2}{\tau+1})-\Li_2(\frac{1-u_3}{\tau+1}) \right)
\\
 + d\log (\tau+u_3) \left( \log \frac{\tau+u_3}{\tau}\log \frac{u_3}{\tau+u_3} + \Li_2(1-u_1) -\Li_2(1-\frac{u_1}{\tau+u_3})\right)
 \\
  + d\log \frac{\tau+u_3}{\tau + v} \left(
     \Li_2(1-\frac{u_1\tau}{\tau+u_3})+\Li_2(1-\frac{u_2}{\tau+1})+\log \frac{u_1\tau}{\tau+u_3}\log\frac{u_2}{\tau+1}-\frac{\pi^2}{6} \right)
  \end{array}\right),
\nl
  I_{2,2} &=& \log\epsilon^2 \times \int_0^\infty d\left( \log \frac{\tau+u_3}{\tau} \log \frac{u_3}{\tau+u_3}
  \right).
\ea The non-spacetime ratio $v=\frac{\l 5624\r\l 6123\r}{\l
5623\r\l 6124\r}$ is needed to produce a parity-odd part.

We emphasize that this comes directly out of the collinear limit of
the heptagon. No manhandling has been applied, nor
would have been necessary. We have, in the interest of this
presentation, used standard dilogarithm identities to simplify the
expression and hopefully make it more human-readable, but we have
not used integration by part nor any manipulation which would affect
the numerical value of the $\tau$-integrand.

The divergent terms cancel upon integration: $I_{1,2}=I_{2,2}=0$,
just as in the pentagon example. This cancelation is of paramount
importance to our approach, and after it is effected, we are left
with two finite and manifestly conformal-invariant integrals
$I_{1,1}$ and $I_{2,1}$. The mechanism for this cancelation is
general and detailed in subsection \ref{sec:nologepsilon}.

The integrals produce trilogarithms. Computing them is not entirely
trivial (for instance, Mathematica would not do them automatically),
but obtaining their symbols is, following, for instance, the method of Appendix~\ref{app:differentials}.
From the symbol it is not too difficult to obtain actual functions, and then fix beyond-the-symbol
ambiguities using the differential computed in Appendix~\ref{app:differentials}.
The resulting functions are quite
simple \ba
  I_{1,1}  &=& \left(\frac13\log^2u_3 +\log u_1\log u_2+ \sum_{i=1}^3 \Li_2(1-u_i)\right)\log u_3 -2\Li_3(1-1/u_3), \nl
  I_{2,1} &=& -\frac12 I_6^{6D} + \Li_3(1-1/u_2) + \Li_3(1-1/u_3)- \Li_3(1-1/u_1) +\frac1{12}\log^3 \frac{u_2 u_3}{u_1}
  \nl&&
  +\frac12\log\frac{u_2 u_3}{u_1} \sum_{i=1}^3 \Li_2(1-1/u_i),
\ea where $I_6^{6D}$ is the six-dimensional massless hexagon
integral~\cite{arXiv:1104.2781,arXiv:1104.2787}, reproduced in
Appendix~\ref{app:functions} alongside the definitions of $x^{\pm}$
and $L_4^+$ to be used shortly.

To complete the computation of $\bar Q R_{6,0}\equiv d R_{6,0}$, we go back to \Eq{Qbar6} and add the other edges contribution via cyclic
symmetry. For future reference, we record the simple result: \ba
 d  R_{6,0}^\textrm{2-loop} &=& I_6^{6D} d \log \frac{ x^+}{x^-} + \left(I_{1,1} d\log \frac{1-u_3}{u_3} + \textrm{two cyclic} \right). \label{dR6}
\ea This agrees precisely with the differential of Goncharov,
Spradlin, Vergu and Volovich's formula~\cite{Goncharov:2010jf},
derived from the results in~\cite{DelDuca:2010zg}, \be
 \frac14 R_{6,0}^\textrm{2-loop} = \sum_{i=1}^3\left(L_4^+(x^+u_i, x^-u_i) - \frac12 \Li_4(1-1/u_i)\right) -\frac18 \left( \sum_{i=1}^3 \Li_2(1-1/u_i)\right)^2
   + \frac{1}{24}J^4 + \frac{\pi^2}{12} J^2 +\frac{\pi^4}{72} \label{R6}.
\ee Of course, in practice the step from \Eq{dR6} to \Eq{R6} can be
a very difficult one, and we do not wish to imply otherwise; we have
simply gone the other way, taking the derivative of \Eq{R6}.
Still, it is impressive how close to \Eq{R6} the present
formalism lands us, namely, on \Eq{dR6}. Important qualitative
features of the result, such as its finiteness, transcendental degree
and conformal invariance, were manifest at every stage of the computation.

\subsection{The differential of the $n$-gon}

To obtain results for $n>6$ two-loop MHV amplitudes, we need the $(n{+}1)>7$-point one-loop NMHV amplitudes. Since there are no qualitative
differences between $(n{+}1)>7$-point amplitudes and the seven-point one, the computation is similar in every respect. We have verified that the
divergent terms integrate to zero for generic $n$, leaving a set of finite and manifestly conformal integrals, which are too lengthy to record
here. However, we have explicitly obtained these integrals for $n=6,7,8,9$ (which is generic) using the present method, and we can compare this
result with that given in~\cite{arXiv:1105.5606} (specifically, equations (4.21) and (4.28) there). We find perfect agreement: numerically both
one-dimensional integrals give the same to 30-digits precision on a few randomly generated Euclidean kinematic points, and symbolically, they
give the same symbol. We recall that these integrals give degree-three transcendental functions characterizing the full differential of the
amplitudes.

\section{Two-loop NMHV and three-loop MHV amplitudes}\label{sec:NMHV}

\subsection{The two-loop NMHV hexagon}\label{sec:NMHVhex}

Because one-loop N${}^2$MHV amplitudes are known, there is no reason
to stop at MHV level. (For one-loop amplitudes, we have used
expressions based on the box-expansion and generalized unitarity
expressed in momentum twistor
space~\cite{bddk2,Cachazo:2008vp,henn08,GrassmannianDCI,Grassmannian2}.)
The first step in our procedure to compute the NMHV hexagon is to
take the collinear limit of the one-loop seven-point N${}^2$MHV
amplitude and extract the $d\epsilon/\epsilon$ term from the
$d^{2|3}\ZZ$ integration. Just as in the previous cases, one obtains
a one-dimensional integral over the variable $\tau$, and after
verifying that terms proportional to $\log \epsilon$ integrate to
zero, one is left with a manifestly finite and conformal integral
over polylogarithms of degree two. The integrals are not
significantly more difficult than those appearing in the MHV case,
and can be done similarly; we only record the result: \ba
\bar Q R^\textrm{2-loop}_{6,1} &=&
 (6) \bar Q\log\frac{\l\bar62\r}{\l\bar64\r} f_1
 + ((1)-(2)+(4)-(5))\bar Q\log\frac{\l\bar64\r}{\l\bar62\r}f_2
 + ((2)-(4))\bar Q\log\frac{\l\bar64\r}{\l\bar62\r}f_3\nl &&
 + \left((6) \bar Q\log\frac{\l\bar62\r}{\l\bar63\r}+((5)-(4))\bar Q\log\frac{\l\bar62\r}{\l\bar64\r}\right)f_4
 + ((2)+(4))\bar Q\log\frac{\l\bar62\r}{\l\bar64\r} f_5\nl &&
 + (5) \bar Q\log\frac{\l\bar62\r}{\l\bar63\r}f_6 + (3) \bar Q\log\frac{\l\bar62\r}{\l\bar63\r}f_7,
\ea where $f_1,\ldots, f_7$ are degree-three transcendental
functions reproduced in Appendix~\ref{app:functions}.

The fact the result could be expanded over a basis of 7 linearly independent rational prefactors (of the form (R-invariant)$\times \bar Q
\log\frac{\l \bar n i\r}{\l \bar n j\r}$), times integrals with unit residues, follows from a general Grassmannian analysis. The key fact is
that these prefactors all originate from seven-point N${}^2$MHV leading singularities, which are combinations of the 15 independent residues in
the $G(2,7)$ momentum twistor Grassmannian~\cite{Grassmannian,Grassmannian2}. In fact, we found that coming up with the full list of the 7
prefactors was the most nontrivial part in our derivation of the above equation. After this was known, the $\mbox{res}_{\epsilon=0}$ part of the
$d^{2|3}\ZZ$
integration step could be easily automated on a computer.%
\footnote{We used a semi-numerical procedure, in which we evaluated
numerically the $d^{2|3}\ZZ$ integral of the N${}^2$MHV residues for
a set of random integer-valued momentum twistors. We then used the analytic
knowledge that the result should be an integral linear combination
of 7 basic objects to promote the numerical result to
an analytic one. Although semi-numerical, this procedure has no
error bars and is rigorously exact.}  There remained only the $\tau$ integration, which could be done automatically
at the level of the symbol and with a bit of human input for the function.

After obtaining this equation, we are (already) essentially done. To
complete this computation, we need to use cyclic symmetry to obtain
the contribution of other edges, and plug the result into the exact
same linear algebra problem as encountered in the one-loop example
in subsection \ref{sec:6pt1loop}.  Namely, starting from the 42
rational prefactors obtained from symmetrizing the above 7 over the
6 edges, we need to add ``zero'' in the form of equation
(\ref{repofzero}) to make the argument of all $\bar Q$'s become
cross-ratios. Just like at one-loop, we found exactly one potential
obstruction, which vanished for the above $f_i$'s, leaving 41 truly
independent functions. We expect this counting to be the same at all
higher loop orders. Expressing the amplitude in the form
\cite{Kosower:2010yk} \ba R^\textrm{2-loop}_{6,1} &=& \frac12\left(
[(1)+(4)]V_3 + [(2)+(5)]V_1 + [(3)+(6)]V_2\right.\nl &&\left.+
[(1)-(4)]\tilde V_3 + [(5)-(2)]\tilde V_1 + [(3)-(6)]\tilde
V_2\right), \label{resnmhv6}\ea
then the solution yields the differentials of each $V$'s and $\tilde
V$'s.  The resulting formulas are reported in Appendix
\ref{app:functions}, together with the definition of $y$ variables.

From these differentials we can already check that the symbols of
$V$ and $\tilde V$ agree with those obtained recently by Dixon,
Drummond and Henn~\cite{Dixon:2011nj}, attached with
their arXiv submission; they do. This is one first nontrivial check.
But we are also interested in beyond-the-symbol information.
We could in principle compute the differential of the results in~\cite{Dixon:2011nj} and
compare with Appendix~\ref{app:functions}, but we have contented
ourselves with a numerical comparison.

To obtain numerical results we first need the value of the functions $V_1,V_2,V_3$ and $\tilde V_1,\tilde V_2, \tilde V_3$ at at least one
point. Using the fact that $V_1+V_2$ and $\tilde V_3$ vanish in the $u_1\to 0$ collinear limit, for instance, we could in principle evaluate
these combinations at any point by integrating along a path connecting to this limit (choosing a path which remains in Euclidean kinematics). We
would then use other paths to compute the other cyclically related combinations. However, we did not find this approach particularly convenient
in practice. A more fruitful strategy is to first derive the amplitude at some other point away from a collinear limit. In fact, in the special
case $u_1=u_2=u_3=u$, it turns out that the differential simplifies dramatically \ba
 dV &=& -I_6^{6D} d\log y +(2\Li_3(1-u)+4\Li_3(1-1/u)+5\log u \Li_2(1-u) + \frac{4}{3}\log^3 u-\frac{4\pi^2}{3}\log u)d\log u
 \nl && -(6\Li_3(1-u)+6\Li_3(1-1/u) +6\log u \Li_2(1-u)+2\log^3 u-2\pi^2\log u)d\log (1-u),
 \nl d\tilde V&=& 0.
\ea
where $V:=V_1=V_2=V_3$ and $\tilde V:=\tilde V_1=\tilde V_2=\tilde V_3=0$, allowing it to be integrated explicitly
\ba
  V(u,u,u)&=& -4L^+(x^+u,x^-u) -\frac{1}{18}J^4-\frac{\pi^2}{9}J^2+2(\Li_4(u)+\frac16\log^3 u\log(1-u))\nl
  && -6\Li_4(1-u)-6\Li_4(1-1/u)+4\Li_3(1-u)\log u-5\Li_2(1-u)\Li_2(1-1/u)
  \nl &&+\frac{7}{24}\log^4u-2\pi^2\Li_2(1-u)-\frac{5\pi^2}{6}\log^2u-2\zeta(3)\log u+\frac{\pi^4}{10}. \label{Vuuu}
\ea The first three terms are essentially as in $-1/3R^{(2)}_{6,0}$.
To fix the constant, we have used numerical integration as explained
in the previous paragraph, connecting these configurations to a
collinear limit. We have computed the value of the constant at the
three points $u=1/3,3/4$ and $5/6$; each point produced the same
result. We then recognized this numerical result as $\pi^4/10$ and
confirmed it to 40 digits. Because it is fully manifest from the
formulation that the constant is a degree four transcendental number
with order one rational coefficient, it does not seem necessary to
supplement the numerics with an analytic computation.

The upshot of the formula is that it is very easy to deform any kinematical point to one on the line $u_1=u_2=u_3=u$.  Integrating the
differential along such paths, we can evaluate efficiently the ratio function at any point. In particular, we have evaluated it on the kinematic
point in~\cite{Kosower:2010yk}. Defining $V', \tilde V'$ by adding the one-loop shift $-\frac{\pi^2}{3}R^{(1)}_{6,1}$ and multiplying by $1/4$
to account for expanding in $a$ as opposed to $2g^2$, we find on the kinematical point
$(u_1,u_2,u_3)=(\frac{112}{85},\frac{28}{17},\frac{16}{5})$ (see \Eq{cr6})
\begin{align}
  V_1' &= 12.6138748750304719319, & \tilde V_1' &= -0.121176561122269858950 i \\
  V_2' &= 11.7057979933899946922, & \tilde V_2' &=  0.030638530205807842307 i\\
  V_3' &= 14.4289552936316184920, & \tilde V_3' &=  0.090538030916462016643 i.
 \end{align}
This was obtained by integrating along a simple linear path in cross-ratio space to the point $u=9/4$, but we have also tried a few other points
and got the same result. The parity even objects $V_i'$ agree precisely with the quantities called $V_i+R_6$ in~\cite{Kosower:2010yk} and the
parity odd objects $\tilde V_i'$ agree within numerical accuracy with those given in \cite{Dixon:2011nj}. After accounting for the same coupling
constant shift, \Eq{Vuuu} can also be compared directly with Eq.(6.30) of \cite{Dixon:2011nj}; we have compared the value at the two points
given in Appendix D of \cite{Dixon:2011nj} and found perfect agreement. Given that their symbols match, these numerical tests remove any doubt
in our mind that the two expressions are equal.

\subsection{The two-loop NMHV heptagon and the three-loop MHV hexagon}

The NMHV heptagon can be attacked in an entirely similar way
starting from the collinear limit of the known one-loop N${}^2$MHV
octagon.

The first step is essentially kinematic and independent of loop
order: one has to list all (rational) objects which can arise from
taking residues of the $d^{2|3}\ZZ_7$ integral on octagon leading
singularities. We found 42 linearly independent ones, all of the
form (R-invariant) times $\bar Q  \log\frac{\l \bar n i\r}{\l \bar n
j\r}$, where $i$ and $j$ are momentum twistors or intersections of
the momentum twistors entering the R-invariants.  An example being
$[23457]\bar Q \log \frac{\l \bar 7 (23)\cap(457)\r}{\l \bar 72\r\l
3457\r}$, but actually only three elements of the basis contained
intersections. In general at $\ell$-loop we expect to find the same
42 structures, each multiplying a pure transcendental integral over
degree $2(\ell-1)$ functions (with potentially a finite a number of additional ones related to 8-point leading singularities not visible at one-loop, which we have not considered). In the case at hands, over
dilogarithms. Another purely kinematic step is the analog of the
linear algebra problem encountered previously: out of the $7\times
42=294$ residues obtained by cyclic symmetry, one has to find all
combinations which can be written $\bar Q\log $ of (conformal
invariant object), possibly adding zero in the form of
\Eq{repofzero}. We found 288 combinations, leaving 6 constraints on
the integrals. These 288 combinations are independent of loop order.

We have not computed the resulting 42 integrals (each of which,
manifestly, would give trilogarithms), but we have computed their
symbol. This was essentially automatic using the method of Appendix
\ref{app:differentials}. Plugging the result into the solution of
the linear algebra problem then gives the symbol of the amplitude.
All entries of the symbol are either four-brackets or intersections
of the type $\l 12(\bar 4)\cap(\bar6)\r$ or $\l 23(745)\cap(\bar 7)\r$.
We hope to analyze it further elsewhere.\footnote{In its present
unprocessed form, the result is too lengthy to be attached with this
arXiv submission. It is available upon request to the authors.}

In this paper, our interest in the heptagon stems mostly from its
connection with the three-loop MHV hexagon via the $\bar Q$
equation. In fact, as already familiar from our analysis of the
two-loop MHV hexagon, in an appropriate basis out of the 15
independent R-invariants at 7-points only five survive
$d^{2|3}\ZZ_7$ integration, namely, $[12367]$, $[12467]$, $[23467]$,
$[23567]$ and $[24567]$. If we write $d
R^\textrm{3-loop}_{6,0}=d\log \frac{\l 5613\r}{\l 5612\r} I_1 + d\log
\frac{\l 5614\r}{\l 5612\r} + \textrm{cyclic}$, it follows that we
can write \ba
 I_1 &=& \int_0^\infty \left(d\log(\tau+1) g_1 + d\log \frac{\tau+1}{\tau} g_4 + d\log \frac{\tau+1}{\tau+u_3} g_3\right)
\nl
 I_2 &=& \int_0^\infty \left(d\log(\tau+v) g_2 + d\log \frac{\tau+v}{\tau} g_5 + d\log \frac{\tau+u_3}{\tau+v} g_3\right) \label{R63}
\ea where $v=\frac{\l 5624\r\l 6123\r}{\l5623\r\l 6124\r}$.  The
five functions $g_i$ are pure degree-four transcendental functions
determined by the collinear limit of the heptagon ratio function. On
physical grounds (the $\tau$ integrals must converge), we know that
$g_{1,2}$ must vanish at $\tau=\infty$ and $g_{4,5}$ must vanish at
$\tau=0$. Thus we can use integration by parts: \ba
 I_1 &=& \int_0^\infty \left(\log(\tau+1) h_1 + \log \frac{\tau+1}{\tau} h_4 + \log \frac{\tau+1}{\tau+u_3} h_3\right)+g_3(0)\log u_3 ,
\nl
 I_2 &=& \int_0^\infty \left(\log(\tau+v) h_2 + \log \frac{\tau+v}{\tau} h_5 + \log \frac{\tau+u_3}{\tau+v} h_3\right)+g_3(0)\log \frac{v}{u_3} .
\ea where $h_i=-\frac{d}{d\tau} g_i$ are degree 3 functions. We see that only the collinear limit of the \emph{differential} of the heptagon is
needed, with the exception of $g_3(\tau=0)$, but one could argue that it is fixed by cyclic and parity symmetry of $d R^\textrm{3-loop}_{6,0}$.
We hope to use these equations in the future to study the differential of the three-loop MHV hexagon, beyond the symbol. In any event, our
result for the symbol of the heptagon already gives the symbols of the $g_i$, which, after the nontrivial but entirely automated integration in
\Eq{R63}, give the symbols of $I_1$ and $I_2$, which in turn give, directly, the symbol of $R_{6,0}^\textrm{3-loop}$. We now describe this
result.

Recently, an Ansatz was constructed for the three-loop hexagon,
based on reasonable physical assumptions about entries of its symbol~\cite{Dixon:2011pw}
(most significantly, that they should all be products of momentum
twistor four-brackets), on OPE constraints
\cite{Alday:2010ku}, and on requiring that that the last entry of
the symbol should involve only brackets of the form $\l
i{-}1ii{+}1j\r$. This Ansatz contained many coefficients but, remarkably,
in the end all but two could be determined by these authors.
Recently Lipatov and collaborators, considering Regge limits using new results on the adjoint representation BFKL
kernel, confirmed the value of a number of these coefficients~\cite{arXiv:1111.0782}.

There are three things we wish to add here.  First, that all entries
of the symbol should be four-brackets is manifest from our approach,
since it follows from the symbol of the two-loop heptagon involving
only momentum twistor intersections (together with the way symbols
of integrals are built using e.g. the algorithm in Appendix
\ref{app:differentials}, and the fact that at six points all
momentum twistor intersections become reducible to four-brackets).
In turn, this property of the heptagon was essentially inherited
from properties of the one-loop octagon in collinear limits. In this
way it should also be possible to obtain general information about
the symbol at $\ell \geq 4$ loops, although we will not do so here.

Second, the assumption about the last entry, conjectured in
\cite{arXiv:1105.5606}, can actually be derived from \Eq{Qbar} and
is therefore now proved. Indeed it follows from writing the NMHV
heptagon in the form (R-invariants) times (pure transcendental
functions), and using the general result for $d^{2|3}\ZZ$ on
R-invariants, equation (\ref{Rinv}). The upshot is that these two
assumptions made in~\cite{Dixon:2011pw} follow rigorously from
\Eq{Qbar}, without doing any explicit computation. Note that
although this form of the heptagon is not strictly proven to all
loops, it is widely believed that it does hold~\cite{Grassmannian},
and assuming this then the statement about the last entry of the
symbol for MHV amplitudes follows to all loops.

Third, we have found that our first-principle computation of the three-loop hexagon symbol is consistent with the Ansatz of \cite{Dixon:2011pw},
which gives a highly nontrivial check on both our approaches. Our new result can be summarized very succinctly: the final two coefficients in
\cite{Dixon:2011pw} are $\alpha_1=-\frac38$ and $\alpha_2=\frac{7}{32}$.

\section{All-loop validity of the $\bar Q$ equation}\label{sec:anomalousdim}

In this section we would like to explain how we believe \Eq{Qbar} could be proved,
and show its consistency at any value of the coupling.

\subsection{Outline of a derivation}

Our proposed derivation of \Eq{Qbar} starts from an expression in~\cite{arXiv:1010.1167,arXiv:1105.5606} for the right-hand side of $\bar Q$ in
terms of insertion of a fermion operator on the edges of the chiral Wilson loop (defined on $(x,\theta)$ space): \be
 \bar Q^{A}_{\dot\alpha} \langle W_{n,k}\rangle \propto
  g^2 \oint dx_{\dot\alpha\alpha} \langle (\psi^A + F\theta^A+ \ldots)^\alpha W_{n,k}\rangle.  \label{QbarW}
\ee
This can be decomposed into a sum of $n$ terms, one for each edge.
Since each edge contribution is gauge invariant and meaningful,
for the following discussion it will suffice to consider the contribution of edge $n$.
For simplicity we will also assume that the (unbroken) $Q$ supersymmetry has been used to set fermions $\chi_{n{-}1}$, $\chi_n$ and $\chi_1$ to zero.
Then $\theta=0$ along that edge,
and the formula reduces to the supersymmetry transformation law of a bosonic Wilson line in a suitable normalization.

The above equation holds for the Poincar\'e supersymmetries of the Wilson loop. For the superconformal generators $S^{\alpha A}$, extra terms are expected due to the breaking of conformal invariance. On the other hand, there is no need to study $S$ explicitly because in the end when we obtain equations for $R_n$, which is known to be conformally invariant.  On $R_n$, the action of $S$ is simply related to that of $\bar Q$.

In~\cite{arXiv:1105.5606}, the chiral Wilson loop with fermion insertion was computed in explicit examples
using conventional Feynman diagram techniques.
The new ingredient in the present paper is a simple yet powerful fact about the spectrum
of excitations of the null Wilson loop: \emph{the fermion
insertion is the unique twist-one excitation with the quantum numbers of $\bar Q$}.

\begin{figure}\centering
\includegraphics[height=4cm]{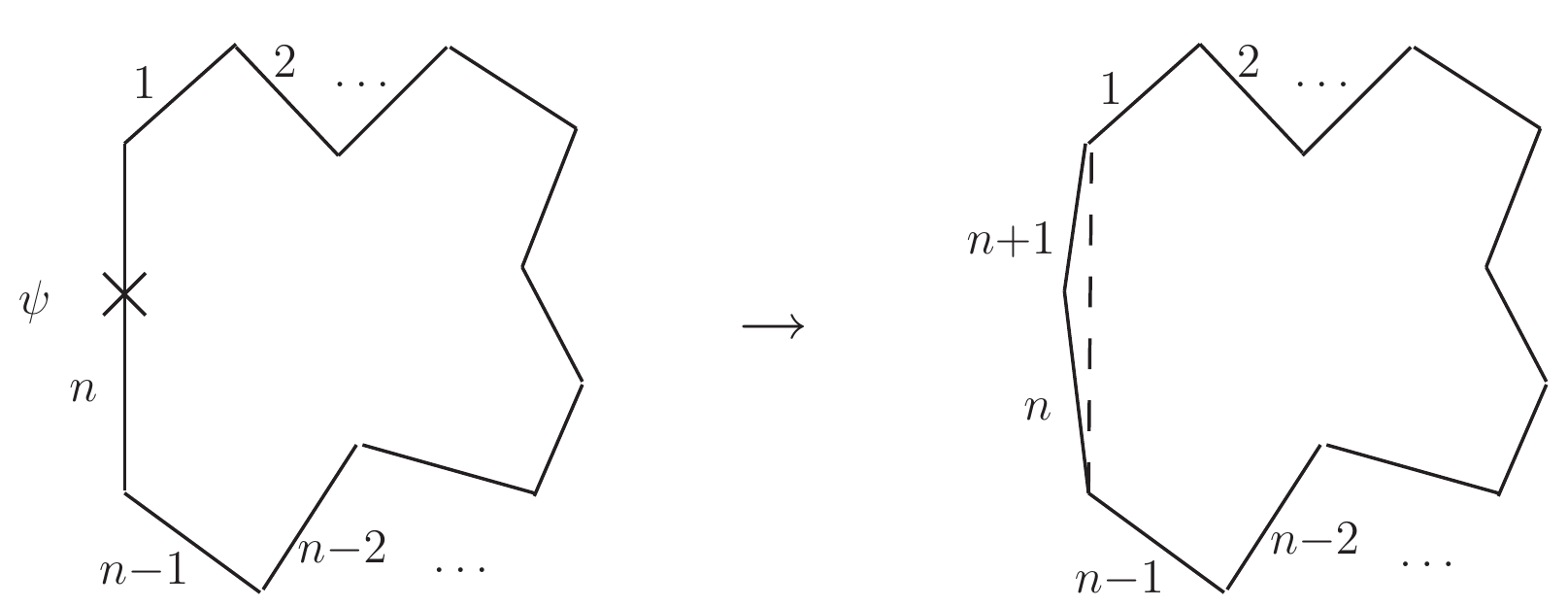}
\caption{Fermion insertion on the Wilson loop versus kink insertion} \label{fig:WL}
\end{figure}

This is a powerful statement because it means that the right-hand side of \Eq{QbarW} isn't really a new object.
Rather, in the spirit of the Operator Product
Expansion (OPE) of null polygonal Wilson loops~\cite{Alday:2010ku},
its expectation value can be extracted from any object
having a nonzero overlap with it in the OPE limit.
The rest of this derivation will thus be based on the analysis of~\cite{Alday:2010ku}.
The simplest possible object is the collinear limit of a $(n{+}1)-$point Wilson loop; this is depicted in Fig.~\ref{fig:WL}.
A good strategy to extract the piece with the right twist and quantum numbers in this limit
is to write down the simplest operation with the quantum numbers of $\bar Q$, namely the $d^{2|3}\ZZ_{n{+}1}$ operation detailed
in section \ref{sec:Qbar}.

To be more precise, the fermion insertion is part of a one-parameter
family of insertions having bare twist one (at weak coupling),
labeled by a position $\tau$ along the edge. In the quantum theory,
operators in this family will renormalize among themselves. Thus we
expect a relation of the form \be
  \lim_{\epsilon\to 0} d^{2|3}\ZZ_{n{+}1}(\tau,\epsilon) \l W_{n{+}1,k}(\tau,\epsilon)\r =  \int_0^\infty d\tau' \tilde F(\tau,\tau',\epsilon) \l \psi(\tau')W_{n,k}\r,
\ee
where the inserted twistor $\ZZ_{n{+}1}(\tau,\epsilon)$ is parametrized as in \Eq{expansion}, and
the right-hand side contains the Wilson loop with insertion we are interested in.
Now we could instead consider the BDS-subtracted Wilson loop, and using the collinear limit properties
of the BDS Ansatz we would find a similar equation with a slightly different $F$
\be
  \lim_{\epsilon\to 0} d^{2|3}\ZZ_{n{+}1}(\tau,\epsilon) R_{n{+}1,k}(\tau,\epsilon) =  \frac{1}{A_n^\textrm{BDS}}\int_0^\infty d\tau' \tilde F(\tau,\tau',\epsilon)  \l \psi(\tau')W_{n,k}\r.
\ee The dependence on $\epsilon$ of the OPE coefficient $\tilde F(\tau,\tau',\epsilon)$ is governed, in the $\epsilon\to 0$ limit, by a
renormalization group equation which we describe in subsection \ref{sec:nologepsilon}. The essential conclusion is that the total integral
$\int_0^\infty d\tau$ does not renormalize, e.g., is $\epsilon$-independent in the limit. Thus, for the total integral, which enters \Eq{QbarW},
$\tilde F\to\tilde F(a)$ depends only on the coupling, allowing us to write \be
 \frac{1}{A_n^\textrm{BDS}} \bar Q \l W_{n,k}\r = \frac{g^2}{F(a)} \int \lim_{\epsilon\to 0} \int_{\tau=0}^{\tau=\infty} d^{2|3}\ZZ_{n{+}1}(\tau,\epsilon) R_{n{+}1,k}(\tau,\epsilon)
+\textrm{cyclic}.  \label{xx1}
\ee

An important subtlety at this point is that the integral $\int_0^\infty d\tau$ is singular due to endpoint divergences. As discussed in the next
subsection, the integral is always at most single-logarithmic divergent, reflecting the behavior expected for $\bar Q$ of the logarithm of an
amplitude.
What this means is that in a given ultraviolet regularization scheme the result will be well-defined, but it may depend on the scheme.%
\footnote{In the momentum space introduced in subsection \ref{sec:nologepsilon}, the anomalous dimension are $\mathcal{O}(p^2)$
while the endpoint divergences produce $1/p$ in the small $p$ limit.  This is why it is still safe to ignore entirely the anomalous dimensions in this discussion.}

A natural way to remove the scheme dependence is to divide by the BDS Ansatz, e.g. push the $1/A_n^\textrm{BDS}$ factor on the left-hand side
inside the $\bar Q$. Since the BDS Ansatz is one-loop exact and proportional to $a$ in the exponent, and since $\bar Q$ is first order in
derivatives, this adds a term \be
 \l W_{n,k}\r \bar Q \frac{1}{A_n^\textrm{BDS}}  = -a R_{n,k} \int \lim_{\epsilon\to 0} \int_{\tau=0}^{\tau=\infty} d^{2|3}\ZZ_{n{+}1}(\tau,\epsilon) R_{n{+}1,1}^\textrm{tree} +\textrm{cyclic}.
 \label{xx2}
\ee Adding the two equations \Eq{xx1} and \Eq{xx2} gives \Eq{Qbar}, up to the yet undermined function of the coupling $F(a)$. In the next
subsection we will determine that $g^2/F(a)=a$, using the known fact that $\bar Q R_{n,k}$ must be finite and scheme-independent to all
loops~\cite{Bern:2005iz}. The point is that some cancelations are required to occur between the two terms.

There are various points in this derivation which may not be fully
rigorous.  For instance, we have assumed that a supersymmetric
regularization of Wilson loop existed, but, as pointed out in
\cite{EKS}, in the only regulator scheme which has been tried so far
it may be necessary to add complicated counterterms to define the
correct operator at the quantum level, changing the explicit form of
the Wilson loop.  This means that our derivation is not based on any
explicitly known regulator. On the other hand, the explicit form of
the operator was not really important for the derivation; we only
really used simple physical properties about the excitation spectrum
of the Wilson loop. Furthermore, in the end, everything is expressed
in terms of finite and regulator-independent quantities. For these
reasons, we believe that this derivation is quite robust.

Following the same steps for Wilson loops in theories with less supersymmetries, one would find that the fermion insertion $\psi$ no longer
appears inside the chiral Wilson loop, so the right-hand side of \Eq{QbarW} would be a genuine new object.  So $\mathcal{N}=4$ SYM is special in
being the only gauge theory in which all elementary fields circulate in a chiral superconnection. Still, one might be able to derive similar
equations in other theories by enlarging the class of Wilson loops to be considered. From our viewpoint, the hallmark of integrability is not
the $\bar Q$ equation itself, because as seen in subsection~\ref{sec:uniqueness} it takes one only ever so far, but the existence of a similar
equation for $Q^{(1)}$, which is known so far only for planar $\mathcal{N}=4$ SYM.

From the scattering amplitude perspective, we expect equations
similar to \Eq{Qbar} to be valid at one-loop in other theories as
well (paralleling the results of~\cite{Beisert:2010gn}). On the
other hand, as noted in Introduction, at higher loops it seems quite
difficult, at least to the authors, to justify the absence of
$1\to3,4,\ldots$ splitting terms in theories that have no local
Wilson loop dual.

\subsection{Convergence of the $\tau$ integral}
\label{sec:notau}

Let us consider the second term in \Eq{Qbar}, the BDS-subtraction
term. Near $\tau=0$ it looks like \be
 -a R_{n,k} \bar Q\log \frac{\l \bar n 2\r}{\l \bar nn{-}2\r} \int_0 \frac{d\tau}{\tau}. \label{BDSsubdiv}
\ee
This is divergent, reflecting the infrared logarithms in the BDS Ansatz.

Poles in the $\tau$-integrand originate from poles in scattering
amplitudes, and this divergence can be traced to poles
$1/\l n{-}2 n{-}1nn{+}1\r$ and $1/\l n{-}1nn{+}11\r$.
Since the amplitude only has poles corresponding to physical channels, these are the only two possible poles which could contribute.
Actually, the second pole blows up in the collinear limit regardless of $\tau$ (see Fig.~\ref{fig:WL}).  A constraint
from the $k$-decreasing collinear limit implies that the coefficient of this pole is $R_{n,k}$, so this contribution cancels out between
this term and a corresponding part in \Eq{BDSsubdiv}, even before we take $\tau\to 0$.
The first pole, $1/\l n{-}2 n{-}1nn{+}1\r$, blows up when we take $\tau\to 0$ but this correspond to a soft limit of the amplitude,
in which its coefficient also reduce to $R_{n,k}$.  The analysis of divergences near $\tau=\infty$ is similar.
We conclude that the finiteness of the $\tau$-integrand, observed empirically in the previous sections, is a general fact
which will remain true at any value of the coupling thanks to the nice collinear and soft limits of BDS-subtracted amplitudes.

Note that this is only true when the relative coefficient between the two terms is chosen as in \Eq{Qbar}.
This is the reason why we believe that $g^2/F(a)=a=\frac14\Gamma_\textrm{cusp}$ in \Eq{xx1} exactly in the coupling.

\subsection{Absence of $\log\epsilon$ divergences}
\label{sec:nologepsilon}

In the main text, we are interested in the zero-momentum component (total $\tau$ integral) of the
difference between the two terms in \Eq{Qbar}.  In this subsection, we will be be interested in the $\tau$-dependence
of just the fist term.  In particular, we wish to understand the $\log\epsilon$ terms which arise in the $\epsilon\to0$ limit
at fixed $\tau$.

From general field theory one might expect these logarithms to be related to the anomalous dimensions of local operators insertions on the
Wilson loop. In the context of null polygonal Wilson loops this was formalized recently~\cite{Alday:2010ku}, and we refer the reader to this
reference for more background. In the case at hand the key feature, just alluded to, is that the only insertions with the correct (bare) twist
and quantum numbers are insertions of single fermions.  The absence of multi-excitation states is a considerable simplification. This means that
all pertinent operators are labeled by one parameter, the position along the edge, so we expect the renormalization group to act as convolution.
Actually, the edge has a symmetry which is a combination of a longitudinal Lorentz boost and a dilatation leaving the position of the two cusps
(and the orientation of neighboring segments) unchanged.  In our variables this is generated by $\tau\to \alpha^{1/2}\tau$. It follows that the
renormalization group equation is diagonalized in momentum space. The upshot is that we expect \be
 \lim_{\epsilon\to 0} \log \left(\int_{\tau=0}^{\tau=\infty} \frac{d^{2|3}\ZZ_{n{+}1}}{d\epsilon/\epsilon} \tau^{\frac{ip}{2}} R_{n{+}1,1}\right) \to \log\epsilon\times (E(p)-1) + C(p)
\ee where on the left-hand side, the $\tau$ integral has been performed but not the $\epsilon$ contour integral; on the right-hand side, the
so-called form factor $C(p)$ (which depends on helicity choices) is finite as $\epsilon\to 0$, and the so-called dispersion relation $E(p)$ has
to match that of an elementary fermion excitation of the null edge (equivalent to excitations of the GKP string \cite{GKP}), known exactly to
all values of the coupling thanks to integrability~\cite{basso} (see also Appendix B of~\cite{arXiv:1010.5009}).

The cancelation of $\log\epsilon$ divergences at zero-momentum is very easy to understand from this formula: the energy $E(0)=1$ is protected
by Goldstone's theorem, the zero-momentum fermion being the Goldstone fermion for the breaking of supersymmetry caused by the Wilson loop background.
The condition $E(0)=1$ is also verified within the integrability framework~\cite{basso}.  This shows that the cancelations observed empirically
in sections \ref{sec:twoloopmhv} and \ref{sec:NMHV} are general and will hold exactly in the coupling.

\subsection{The fermion dispersion relation}
\label{sec:dispersion}

We now wish to check the prediction for $E(p)$ at finite $p$.  Due
to the physical origin of the divergences, it should suffice to
check this for the collinear limit of six-point amplitudes, the
dispersion relation being expected to be universal. We let \be
 \lim_{\epsilon\to 0} \frac{\epsilon}{d\epsilon} d^2Z_{n{+}1} \int d^{0|3}\chi_{n{+}1} R_{n{+}1,1}^\textrm{$\ell$-loop}:= \bar Q\log \frac{\l 4512\r }{\l 4513\r} \times \frac{d\tau}{\tau}\times I^\textrm{$\ell$-loop}(\tau).
\ee as in subsection \ref{sec:pentagon}, where $\ell$ is the loop
order (e.g., the order in $a=\frac14\Gamma_\textrm{cusp}$), and
$n=5$.

Taking the collinear limit of the six-point tree amplitude
$(1)+(3)+(5)$ (the rules in subsection \ref{sec:pentagon} can be
useful here) gives \be
 I^\textrm{tree}(\tau)= \frac{1}{\tau+1}.
\ee
We will need the Fourier transform
\be
 I^\textrm{tree}(p) = \int_{-\infty}^\infty \frac{d\sigma e^{\frac{ip}{2}\sigma}}{e^\sigma+1} = \frac{\pi}{i\sinh \frac{\pi p}{2}},
\ee
where $\tau=e^\sigma$.  The specific form of this result is very useful, as it allows us to write the ratio
\ba
 \frac{I^\textrm{$\ell$-loop}(p)}{I^\textrm{tree}(p)} &=& e^{\frac{\pi p}{2}} \frac{e^{-\pi p}-1}{2\pi i} \int_{-\infty}^\infty d\sigma e^{\frac{ip}{2}\sigma} I^\textrm{$\ell$-loop}(e^\sigma)
= \frac{e^{\frac{\pi p}{2}}}{2\pi i} \oint_{\mathcal{C}} d\sigma
e^{\frac{ip}{2}\sigma} I^\textrm{$\ell$-loop}(e^\sigma),
\label{contourform} \ea where $\mathcal{C}$ is the rectangle
contour: \be
\includegraphics[height=2cm]{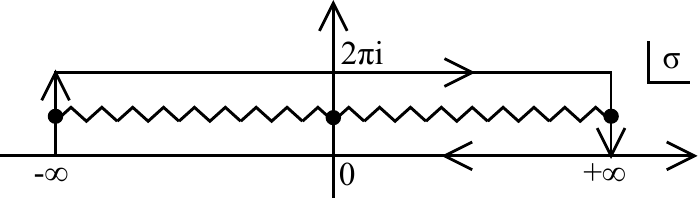}\nonumber
\ee This is valid for $-2<\textrm{Im} p<0$, where the contributions
from infinity can be neglected. Now, for any $\ell$,
$I^\textrm{$\ell$-loop}(\tau)$ is an analytic function of $\tau$
with branch points at $\tau=0,-1$, and $\infty$. This allows the
contour to be deformed and expressed in terms of discontinuities on
the horizontal line $\textrm{Im } \sigma=\pi$:
\ba
  \frac{I^\textrm{$\ell$-loop}(p)}{I^\textrm{tree}(p)} &=& \int_{-\infty}^\infty d\sigma e^{\frac{ip}{2} \sigma} \textrm{Disc}\left[ I^\textrm{$\ell$-loop}(-e^\sigma)\right]
  \nl &=& \int_0^1 \frac{dx}{x} x^{\frac{ip}{2}} \textrm{Disc}\left[ I^\textrm{$\ell$-loop}(-x)\right] + (p\to -p) \label{discform}
\ea where $\textrm{Disc} \left[I^\textrm{$\ell$-loop}(-x)\right]:= \frac{I^\textrm{$\ell$-loop}(-x - i0)-I^\textrm{$\ell$-loop}(-x + i0)}{2\pi
i}$. It could be very interesting to interpret this discontinuity (in the cross-ratio regime $u_1\to 0$, $u_2=1-u_3$, $u_3<0$) as the imaginary
part of the six-gluon amplitude in a physical Minkowski-signature regime.

To deal with such integrals, we found useful to think of
$\frac{ip}{2}$ as a positive integer and use the language of
harmonic sums.  Recalling the 5-point $\tau$ integrand \Eq{5ptint}
and taking its discontinuity, we obtain \ba
\frac{I^\textrm{1-loop}(p)}{I^\textrm{tree}(p)} &=& \left( M\left[
\frac{2x}{(x-1)_+}\right]\log\epsilon + M\left[
\frac{x+1}{(x-1)_+}\log(1-x)\right]-\frac{\pi^2}{6}\right)+(N\to-N)\nl
 &=& \left(2S_1\log \epsilon +\frac{S_1}{N}-S_1^2-S_2 -\frac{\pi^2}{6} \right) + (N\to-N)  \label{I1loop} \label{1loopdiv}
\ea
where $M[f]:=\int_0^1 \frac{dx}{x} x^{N} f$ is the Mellin transform, $N:=\frac{ip}{2}$, and the $+$ prescription is the usual one, such that
$\int_0^1 \frac{dx \log(1-x)^a}{(1-x)_+}=0$ for $a\geq 0$. For integer $N$ the harmonic sums are defined as $S_i=\sum_{n=1}^N \frac{1}{n^i}$
and $S_{i_1,i_2}=\sum_{n\geq n_1\geq n_2} \frac{1}{n_1^{i_1} n_2^{i_2}}$, and elsewhere by analytic continuation \cite{remiddi}.  The $-\pi^2/6$ term
follows from a careful treatment of the $x$ near 1 region, but can also be verified numerically quite unambiguously using the first form in \Eq{contourform}.

At two-loops, we are interested only in the $\log \epsilon$-terms. Conveniently, these can be read off from our formula for the differential of the NMHV hexagon,
in Appendix \ref{app:functions}, without doing any integration.  The point is that the logarithms exclusively arise from terms proportional to $d\log u_1$.
So, all we have to do, is take the expressions in Appendix \ref{app:functions}, drop all terms except those proportional
to $d\log u_1$, and expand the degree three functions in powers of $\log u_1$ as $u_1\to 0$.
Then we use the simple rules $\log u_1 d\log u_1\to 2 \log^2\epsilon, d\log u_1\to 2\log\epsilon$.
We then have to take a discontinuity as a function of $\tau\to-x$. We obtain for the $\log^2\epsilon$ terms
\ba
 \frac14 \frac{I^\textrm{2-loop}}{I^\textrm{tree}} &=& \log^2\epsilon \left( M\left[ \frac{(1{+}x)\log(1-x) -\frac12\log x}{(1{-}x)_+}\right] + \frac{\pi^2}{12} \right)+ (N\to -N)
 \nl &=& \log^2\epsilon \left( S_1^2+\frac12S_2 -\frac{S_1}{N} +\frac{\pi^2}{6}\right)+(N\to -N),
\ea
and for the single-logarithmic terms
\ba
 \frac14 \frac{I^\textrm{2-loop}}{I^\textrm{tree}} &=&  \log\epsilon \left(M\left[ \frac{x(\frac{\pi^2}{6}+\Li_2(x)) + \frac12(1+x)\log(1-x)(3\log(1-x)-\log x)}{(1-x)_+} \right]+\frac12 \zeta(3)\right) + (N\to -N)
 \nl &=&  \log\epsilon \left( -S_{1,2}-S_1S_2-S_1^3+\frac{S_2}{N}+\frac32\frac{S_1^2}{N}-\frac{S_1}{2N^2}-\frac{\pi^2}{2}S_1-\frac52\zeta(3)\right)+(N\to -N).
\ea
The OPE prediction concerns the logarithm, so we need to add the combination
\ba
 - \frac18 \left(I^\textrm{1-loop}(p)\right)^2 &=& + \log\epsilon \left( 2S_{1,2}-S_3+S_1^3-\frac{S_2}{N}-\frac32\frac{S_1^2}{N}+\frac{S_1}{N^2}+\frac{\pi^2}{2}S_1+3\zeta(3)\right) + (N\to -N)\nl
 && -\log^2\epsilon \left(S_1^2+\frac12 S_2 -\frac{S_1}{N} +\frac{\pi^2}{6} \right)+ (N\to -N).
 \ea
This is the square of \Eq{1loopdiv}, although writing it as harmonic sums
was not entirely trivial due to cross-terms between the $+N$ and
$-N$ terms.  We found that an efficient way to achieve this was to match the poles
on the negative $N$ axis. The $\log^2\epsilon^2$ terms cancel in the sum, and we obtain for the
logarithm \be
 \frac14 \left(\log I(p)\right)^\textrm{2-loop} = \log \epsilon \left( S_{1,2}-S_3-S_1S_2 +\frac12\frac{S_1}{N^2}+\frac12\zeta(3)\right) +(N\to -N)+\mbox{finite}.
\ee

This is to be compared with $(E(p)-1)\log \epsilon$ where for $E(p)$
we use the expansion to second order in $\Gamma_\textrm{cusp}$ of
the ``large fermion'' dispersion relation in Eq.(20) from
\cite{basso}: \ba
 E(p)-1 &=& \Gamma_\textrm{cusp} \left(\psi_+ -\psi(1)\right) - \frac{\Gamma_\textrm{cusp}^2}{8}
 \left(\psi_+''  + 4\psi_-'(\psi_--\frac{1}{p}) +6\zeta(3) \right)
 \nl &=& \frac12\Gamma_\textrm{cusp} S_1 + \frac1{4}\Gamma_\textrm{cusp}^2(S_{1,2} -S_3-S_2S_1+\frac{S_1}{2N}+\frac12\zeta(3)) +(N\to-N)\nonumber
\ea
where $\psi_+:= \frac12(\psi(1+\frac{ip}{2})+ \psi(1-\frac{ip}{2}))$ and $\psi_-:= \frac i2(\psi(1+\frac{ip}{2})- \psi(1-\frac{ip}{2}))$ in the first line,
and on the second line we have converted the result to harmonic sums.  Perhaps we should have said earlier, that
the one-loop prediction $2S_1$ was matched by \Eq{1loopdiv}.

The perfect agreement confirms beautifully that our $d^{2|3}\ZZ$
integral is probing fermion excitations on the edges of the Wilson
loop, as was expected from the OPE analysis. Second, and perhaps
more importantly for us, it gives an independent confirmation to
two-loop accuracy (besides the numerical check in subsection
\ref{sec:NMHVhex}), that the prefactor in \Eq{Qbar} has to be
$\Gamma_\textrm{cusp}$.

\section{Two-dimensional kinematics}\label{sec:2d}

\subsection{Preliminaries}

It can be useful to consider special kinematic configurations in
which scattering amplitudes/Wilson loops usually simplify.
Following~\cite{Alday:2009ga,Alday:2009yn} at strong coupling, and
~\cite{arXiv:1006.4127, khoze} at weak coupling,  we now consider
configurations of external momenta/edges of the Wilson loop which
can be embedded inside a two-dimensional subspace of Minkowski
space.  This reduces the conformal group SU(2,2) to
SL(2)$\times$SL(2). Actually, we are interested in super-amplitudes,
and as will become apparent soon it is very natural to consider a
supersymmetric reduction to SU(2,2$\mid$4) to
SL(2$\mid$2)$\times$SL(2$\mid$2).

With no loss of generality we take the number of particles/edges to be even, say $2n$. Even and odd labels are distinguished, \be
\ZZ_{2i{-}1}=(\lambda^{1+}_{i},0,\lambda^{2+}_{i},0,\chi^{1+}_{i},0,\chi^{2+}_{i},0), \qquad \ZZ_{2i}=(0,\lambda^{1-}_i,0,\lambda^{2-}_i,
0,\chi^{1-}_{i},0,\chi^{2-}_{i}),\ee where $i=1,\ldots,n$ for even and odd labels. Four-brackets with two odd and two even labels factorize, $\l
2i{-}1\,2j{-}1\,2k\,2l\r:=\l ij\r [kl]$, and all others vanish.  R-invariants which contain a generic reference twistor $Z_*$, two odd and two
even labels will appear, and they factorize into odd and even parts,$[*\,2i{-}1\,2j{-}1\,2k\,2l]=(*\,i\,j)[*\,k\,l]$ where \be
(*\,i\,j):=\frac{\delta^{0|2}(\l\l*\,i\,j\r\r)}{\l
*\,i\r \l i\, j\r \l j\,*\r} \ee and similarly for the even part
$[*\,k\,l]$. These R-invariants satisfy a four-term identity $(a\,b\,c)-(a\,b\,d)+(a\,c\,d)-(b\,c\,d)=0$. Finally, cross-ratios are also
separated into odd and even sectors \be u^+_{i,j}=\frac{\l ij{+}1\r\l i{+}1j\r}{\l ij\r\l i{+}1j{+}1\r},\quad
u^-_{i,j}=\frac{[ij{+}1][i{+}1j]}{[ij][i{+}1j{+}1]}.\ee In this notation, the NMHV tree amplitude (most easily extracted from the CSW form
\cite{Bullimore:2010pj}) is \be
 R_{2n,1}^\textrm{tree} = \frac12\sum_{i,j} (*\,i\,j) \left( [i\,j{-}1\,j]-[i{-}1\,j{-}1j]\right).
\ee

\subsection{Collinear limits}

The BDS-subtracted amplitudes behave simply under single-collinear limits. In two-dimensional kinematics, the natural limit instead collapses
the length of an edge to zero, the so-called triple-soft-collinear limit. The behavior of amplitudes in this limit is well understood at both
two-loops and at strong coupling~\cite{arXiv:1006.4127,khoze}; let us recall the main conclusion. First, it is easy to see that nothing can
diverge in this limit, just by dual conformal symmetry, to any loop order: for the hexagon the whole (BDS-subtracted) amplitude is just a
constant since all cross-ratios are equal to 1. On the other hand, as is especially clear from the Wilson loop viewpoint where the limit has an
OPE interpretation~\cite{Alday:2010ku,arXiv:1010.5009}, the limit involves only physics localized around one edge and so cannot depend on the
number of points. So in general we expect a simple multiplicative relation $R_{2n}\to f(\aa)R_{2n-2}$ for some function of the coupling.

Actually, we will need both $k$-preserving or $k$-decreasing limits, which are not parity conjugate to each other
in two-dimensional kinematics. This allows for two different constants.
It will be useful to absorb them by a simple rescaling of the BDS-subtracted amplitudes:
\be
 R_{2n,k} := e^{(n-2) f_1(\aa) + k f_2(\aa)} \tilde R_{2n,k}.
\ee
By choosing $f_1$ and $f_2$ such that $\tilde R_{6,0}=1$ and $\tilde R_{6,1}=R^\textrm{tree}_{6,1}$, then
$\tilde R_{2n,k}\to \tilde R_{2n-2,k}$ in the $k$-preserving limit $\lambda_{n{+}1}^-\to\lambda_n^-$, and
$(\int d^2\chi^+_n d^2\chi^-_n \tilde R_{2n,k})/( \int d^2\chi^+_n d^2\chi^-_n (n{-}1n1)[n{-}1n1])\to \tilde R_{2n{-}2,k{-}1}$
in the $k$-decreasing case.  It is known that
\be
 f_1(\aa) = -\aa^2\frac{\pi^4}{9} +\mathcal{O}(a^3), \quad  f_2(a)=-\aa\frac{\pi^2}{3}+\aa^2\frac{7\pi^4}{30} + \mathcal{O}(a^3).
\ee The strong coupling limits are also known: $f_1(\aa)=
\aa\frac{2\pi}{3}+\mathcal{O}(1)$, $f_2(\aa)= 0 + \mathcal{O}(1)$,
when $\aa\to \frac{\sqrt{g^2_\textrm{YN}N_c}}{8\pi}\to\infty$, as we
extract from~\cite{Alday:2009dv}. The result for $f_1$ at weak
coupling has been known for a while~\cite{arXiv:1006.4127}, while
the two-loop correction to $f_2$ follows from the value
$\frac{8\pi^4}{45}A^\textrm{tree}$ of the two-loop value of
$R_{6,1}$ in \Eq{Vuuu}, also obtained recently by
\cite{Dixon:2011nj}. It would be nice to have a way to calculate
these functions at all values of the coupling. Below we will discuss
the functions $\tilde R_{2n,k}$ directly.

\subsection{Two-dimensional Yangian equations}

\Eq{Qbar} involves a $(2n{+}1)$-gon which is not very natural in two-dimensional kinematics. However, the key ingredient in its derivation was
the fact that the collinear limit of the $(2n{+}1)$-gon had a nonzero overlap with the insertion of a fermion on the $2n$-gon.  This suggests
that we can get the same information out of the $(2n{+}2)$-gon.  Namely, we write down a limit with the right quantum numbers. Take $A$ and $a$
to be even and add particles $2n{+}1$ and $2n{+}2$: \ba \hspace{-0.3cm}
 \bar Q^A_a \tilde R_{2n,k}  &=& \aa \int d^{1|2}\lambda^+_{n{+}1}\int d^{0|1}\lambda^-_{n{+}1} (\tilde R_{2n{+}2,k{+}1} - R^\textrm{tree}_{2n{+}2,1}\tilde R_{2n,k}) + \textrm{cyclic}
 \nl &:=& \aa~ \lambda^-_{n\,a} \lim_{\lambda^-_{n{+}1}\to \lambda^-_n} \int_{\lambda_n^+}^{\lambda_1^+} \l \lambda^+_{n{+}1} d\lambda^+_{n{+}1}\r\int d^2\chi_{n{+}1}^+(d\chi_{n{+}1}^-)^A
 \left(\mbox{parenthesis}\right) + \textrm{cyclic}.\quad \label{Qbar2d} \ea In principle, one might expect a nontrivial function of the coupling multiplying the first term, arising as an OPE
coefficient similar to $g^2/F(a)$ above, but not in front of the second, it being associated with the BDS Ansatz. However that function would
have to be independent of $n$, and arguing as follows it is possible to see that using $\tilde R$ this function can only be 1. The essential
point is that some cancelations have to occur between the two terms.

On R-invariants not depending $n{+}1$, the integral gives zero, while in general
\be
 d^{1|2}\lambda^+_{n{+}1} d^{0|1}\lambda^-_{n{+}1} (n{+}1\,i\,j)[n{+}1\,k\,l] =
 \bar Q \log \frac{[n k]}{[nl]}
 d\log \frac{\l \lambda_{n{+}1}^+ \lambda^+_i\r}{\l \lambda_{n{+}1}^+ \lambda^+_j\r}.
 \label{Rinv2d}
\ee The limit in \Eq{Qbar2d} does not depend on how
$\lambda^-_{n{+}1}$ approaches $\lambda^-_n$, but for individual
terms it may, thus it is useful to choose
$\lambda_{n{+}1}^-=\lambda^-_n + \epsilon \lambda^-_1$
supersymmetrically.  Then the previous equation is valid in all
cases, with the substitution $[nn]\to [n1]$ when $k=n$ or $l=n$. The
action on NMHV tree gives \be
  d^{1|2}\lambda^+_{n{+}1}d^{0|1}\lambda^-_{n{+}1} \tilde R_{2n{+}2,1}^\textrm{tree} = \sum_{j=2}^n d\log \l \lambda^+_{n{+}1} \lambda^+_j \r \bar Q \log\frac{[nj{-}1]}{[nj]}.
\ee This agrees, for the generic term, with Eq.(3.13) of \cite{khoze} for the BDS amplitude, but we see that the measure is logarithmically
divergent at the endpoint $\lambda^+_{n{+}1}=\lambda^+_n$.  On the other hand, thanks to the universal $k$-decreasing collinear limits of
$\tilde R$, this divergence will cancel in the combination \Eq{Qbar2d} as in subsection \ref{sec:notau}, showing that it is the correct
combination.
Identical equations apply when both indices on $\bar Q$ are in the odd (plus) sector, if we take
$\lambda_i^+\to\lambda_i^-,\lambda_i^-\to\lambda_{i{+}1}^+$.

The $Q^{(1)}$ equation is trickier, since the two-dimensional subsector is not closed under the action of the naive generator, $Q^{(1)}$. We
hope to address this issue, and make use of both $\bar Q$ and $Q^{(1)}$ equations in restricted kinematics in the future.

\subsection{From tree N${}^2$MHV to one-loop NMHV to two-loop MHV}

Let us now check whether it is indeed possible to compute the
two-loop amplitude in restricted kinematics, starting with tree
amplitudes in \emph{only} restricted kinematics. For the N${}^2$MHV
tree amplitude we start from the CSW
representation~\cite{Bullimore:2010pj}, which has a simple
two-dimensional limit for the generic term, albeit the bondary terms
are a bit complicated. After some massaging we managed to obtain the
form: \ba
 R_{2n,2}^\textrm{tree} &=& \frac12 \sum_{i<j<k<l<i} (*\,i\,j)(*\,k\,l)\left( [i\,j{-}1\,j]-[i{-}1\,j{-}1j]\right)\left( [k\,l{-}1\,l]-[k{-}1\,l{-}1l]\right)
\nl && - \frac13 \sum_{i<j<l<i} (*\,i\,j)(*\,j\,l) \left(
([l\,i{-}1\,i]-[j{-}1\,i{-}1\,i])([j\,l{-}1\,l]-[j{-}1\,l{-}1\,l])
\right. \nl && \left.\hspace{3.5cm}+
([j\,i{-}1\,i]-[j{-}1\,i{-}1\,i])([j\,l{-}1\,l]-[i{-}1\,l{-}1\,l])\right).
\label{n2mhv} \ea

Computing the $\bar Q$ from edge $2n$ using \Eq{Qbar2d} and
\Eq{n2mhv} is a bit tedious, possibly because this form for the
trees is not optimally simple. After patient bookkeeping we obtain
the simple result \be
 \bar Q \tilde R_{2n,1}^\textrm{1-loop} = \!\!\!\!\!\!\sum_{1\leq j<k<l\leq n} \log\frac{\l 1k\r}{\l nk\r} \bar Q\log \frac{[nk]}{[nk{-}1]} (jkl)( [j\,l{-}1\,l]-[j{-}1\,l{-}1\,l] )
+\textrm{cyclic}. \ee To integrate the $\bar Q$ we need to complete
its arguments into cross-ratios.  A successful strategy is to decompose
the logarithms as a difference of two terms and collect the terms.
Then for the generic term we immediately get
cross-ratios, but we also get some boundary terms
\ba
 \bar Q \tilde R_{2n,1}^\textrm{1-loop} &=& \!\!\!\!\!\!\sum_{i<j<k<l<i} \log \l ik\r (\bar Q \log u_{i{-}1,k{-}1}^-)  (jkl)([j\,l{-}1\,l]-[j{-}1\,l{-}1\,l])
- \!\!\!\!\sum_{i<j<k<i} \log \l i k\r (i j k) \nl
 &&  \hspace{-0.5cm}\times \left( \bar Q\log\frac{[i k{-}1]}{[i k]}( [i\,j{-}1\,j]-[i{-}1\,j{-}1\,j]) + \bar Q \log\frac{[i{-}1 k{-}1]}{[i k{-}1]} ([k\,j{-}1\,j]-[k{-}1\,j{-}1\,j])\right).\nonumber
\ea
To complete the arguments of $\bar Q$ in the boundary terms one can in principle follow the systematic strategy of subsection \ref{sec:6pt1loop}.
However in this case this is not necessary, as close inspection reveals that
adding zero in the form $\bar Q \log \frac{[ij]}{[jk]}[i\,j\,k] = \bar Q \log \frac{[ij]}{[jk]} ([k\,j{-}1\,j]-[i\,j{-}1\,j]+[i\,j{-}1\,k])$, antisymmetrized in $(i,i{-}1)$ and $(k,k{-}1)$, will do the trick.
Thus, without much attempt at simplifying the result, we obtain the formula
\ba
 \tilde R_{2n,1}^\textrm{1-loop} &=& \!\!\!\!\!\!\!\sum_{i<j<k<l<i} (jkl)([j\,l{-}1\,l]-[j{-}1\,l{-}1\,l])\log \l ik\r \log u_{i{-}1,k{-}1}^-
-\!\!\!\! \sum_{i<j<k<i} \log \l i k\r (i j k) \nl
 &&  \hspace{-0.5cm}\times \left[ \log u^-_{i,k{-}1,k,j}( [i\,j{-}1\,j]-[i{-}1\,j{-}1\,j]) + \log u^-_{k{-}1,i{-}1,i,j}([k\,j{-}1\,j]-[k{-}1\,j{-}1\,j]) \right.
 \nl && + \left.\left( \big(\log u^-_{i,j,j{-}1,k} [i\,j{-}1\,k] - (i\leftrightarrow i{-}1)\big)-(k\leftrightarrow k{-}1)\right) \right].
\ea
This is the general one-loop NMHV amplitude in two-dimensional kinematics.\footnote{The
first arXiv submission of this paper contained a typographic mistake, with $(k\leftrightarrow k{-}1)$ appearing as $(j\leftrightarrow j{-}1)$ in the last line.}
We remark that the formula naively contains ill-defined $\log [jj]$ terms from $k=j{+}1$ in the last sum, however these cancel within the square bracket making the formula well-defined.  Furthermore the formula respects conformal invariance, although it is not expressed explicitly in terms of cross-rations.
We have verified that this formula agrees with the two-dimension limit of the box expansion for $2n=8,10,12$.

To get the two-loop MHV amplitudes from this using the $\bar Q$
equation is much simpler, as \Eq{Rinv2d} turns out to reduce to the substitution
$(*n{+}1j)[*n{+}1k]\to \log \frac{\l 1j\r}{\l nj\r}\log [nk]$ in this case; this yields \ba
R^{\textrm{2-loop}}_{2n,0}&=& -\sum_{i<j<k<l<i} \log \l i k \r\log
\l j l\r \log u^-_{i{-}1,k{-}1} \log
u^-_{j{-}1,l{-}1}-2\sum_{i<j<k<i}\log \l i j \r\log \l j k\r \nl
&&\hspace{-1cm}\times (\log u^-_{j{-}1,k{-}1} \log
u^-_{k{-}1,i,i{-}1,j} + \log u^-_{i{-}1,j{-}1} \log
u^-_{j{-}1,k,k{-}1,i} + \log u^-_{i{-}1,k{-}1} \log u^-_{l,
j,j{-}1,k{-}1}) \nl
 && -\sum_{i<j<i} \log^2\l ij\r \log u^-_{i{-}1,j{-}1}\log(1- u^-_{i{-}1,j{-}1}).
\ea where in the parenthesis the second and third terms are
permutations of the first one. We have verified that this result is
equivalent to the forms given in \cite{khoze,arXiv:1010.5009},
albeit its form is not immediately as elegant.

Something unusual has to be said.  We found that obtaining a reasonably intelligible form of the N${}^2$MHV tree amplitude in restricted
kinematics was by far the most time-consuming part of this computation. Obtaining the NMHV one-loop result from it, involved some patient
bookkeeping, but no special difficulty.  Finally, obtaining the two-loop MHV formula was by far the simplest step. While we hope that simpler
expressions for the tree and one-loop amplitudes will be obtained in the future, our main objective here was to confirm that it was possible to
obtain these results using only input from two-dimensional amplitudes.

\section{Conclusion}

We have proposed an all-loop equation for the $\bar Q$ symmetry
acting on BDS-subtracted planar S-matrix of $\mathcal{N}=4$ SYM, as
well as its parity-conjugate for $Q^{(1)}$ symmetry, which,
interpreted as quantum-corrected symmetry generators, amounts to
exact Yangian symmetry. In principle, these equations can be used to
determine all-loop S-matrix uniquely, at at any value of the
coupling. As a perturbative study, we have applied the $\bar Q$
equation to reproduce results for MHV $n$-gon, NMHV hexagon at two
loops, and fix the undetermined coefficients in an Ansatz for
three-loop MHV hexagon. The equations relate dual Wilson loops with
fermion insertions along edges to higher-point ones in the collinear
limit, and we have outlined a derivation in the spirit of OPE. In
particular, we have reproduced the fermion dispersion relations to
the second order of $\Gamma_{\textrm{cusp}}$, as a strong
consistency check for the proposal.

Our main finding is that scattering amplitudes are precisely Yangian invariant,
provided that the correct quantum mechanical expressions for the generator are used, and are fully determined by that.

It should be possible to extend our results in the more loops/more legs directions, with a manageable amount of time. For instance, higher-point
three-loop MHV amplitudes could be analyzed. Using parity the N${}^2$MHV heptagon is equivalent to the NMHV heptagon (whose symbol we have been
able to compute at two loops, though not its function), and from it the three-loop NMHV hexagon could be obtained. The two-loop N${}^2$MHV
octagon should be uniquely determined using parity symmetry together with the $\bar Q$ equation; from this one could obtain all three-loop
heptagons and four-loop hexagons.  With, perhaps, the help of a big enough computer, we foresee no essential difficulty in obtaining the symbol
of these objects. Without doing detailed computations, it might also be possible to make general qualitative statements about these objects.

In addition to two- and three-loop results, we have also obtained certain all-loop predictions for prefactors multiplying degree-$(2\ell-1)$
functions for the $\bar Q$ of $\ell$-loop NMHV hexagon and heptagon, which give a list of the last entry of their symbols. Based on
classifications of residues in the Grassmannian $G(2,n)$, it should be straightforward to generalize such predictions to NMHV $n$-gon, in analog
of that of MHV $n$-gon. More importantly, by considering the $Q^{(1)}$ equation, we hope to carry such analysis for N${}^k$MHV cases, which
would provide useful information on the all-loop structure of general amplitudes.

Corresponding functions, as opposed to symbols, could also be obtained provided one is able to carry out the one-dimensional integrals which
appear, but this might require some new ideas. Clearly it would be important to better understand results which are already known, for instance
the differential of the general MHV $n$-gon~\cite{arXiv:1105.5606} and the three-loop MHV hexagon. Indeed the undetermined coefficients in the
Ansatz of~\cite{Dixon:2011pw} at three-loops are now completely fixed, and it would be fascinating to see the corresponding function.

It would also be fascinating to study our equations at strong
coupling, paralleling the the strong coupling application of OPE.
It is consistent with all which is presently known at strong coupling, in that it express $\bar Q \log R_{n,0}$
as $\sqrt{\lambda}$ times a ratio function, which is known to be of order 1.  However, to test it in a nontrivial way,
one would have to obtain the ratio function, which corresponds to a one-loop computation in the string theory.

Amplitudes restricted to a two-dimensional subspace of four-dimensions seem especially promising. 
There is the intriguing possibility, that the main complexities of loop amplitudes in these kinematics might already be present in some way
inside the tree amplitudes, where the combinatorics have been worked out in~\cite{arXiv:1010.6256}. With some effort, one should be able to
obtain all three-loop MHV amplitudes from some nice form of tree N${}^3$MHV. The simplest case would be to derive and fix the coefficients of
the recent conjecture for the octagon~\cite{Heslop:2011hv} from tree
N${}^3$MHV tetradecagon. 

\acknowledgments{SCH would like to thank Nima Arkani-Hamed, Johannes Henn and Tim Goddard for useful discussions. SH would like to thank Niklas
Beisert, Tristan McLoughlin, Jan Plefka and Cristian Vergu for interesting discussions. SCH and SH wish to thank the Mathematica school for
increasing their computer literacy, and acknowledge hospitality at Perimeter Institute and Niels Bohr Institute where this work was started. SCH
gratefully acknowledges support from the Marvin L.~Goldberger Membership and from the National Science Foundation under grant PHY-0969448. SH's
stay at Perimeter Institute was partially funded by the EC ``Unify'' grant.}

\begin{appendix}

\pagebreak
\section{Taking differentials of one-dimensional integrals}
\label{app:differentials}

Consider an integral of the form
\be
  \int_0^\infty d\log (x+a) F(x,u_i),
\ee
or, more generally, a linear combination of such integrals such that the total integrand converges both at zero and infinity.

Then its differential is the sum of the following terms:
\begin{itemize}
 \item A term
\be
-F(x{=}0,u_i) d\log a
\ee
 \item For each term of the form $G_j(x,u_i) d\log (x+x_j)$ in the differential of $F$, a term
 \be +(d\log (a-x_j)) \int_0^\infty (d\log \frac{x+a}{x+x_j}) G_j(x,u_i)
 \ee
\item For each term of the form $H_j(x,u_j)d\log f$ in the differential of $F$, where $f$ is independent of $x$, a term
\be + (d\log f) \int_0^\infty d\log (x-a) G_0(x,u_i).
\ee
\end{itemize}
The proof is left to the reader; it is more or less integration by parts.
Sometimes it may happen that some $x_j=0$, in which case some intermediate expressions will be ill-defined.
This can be dealt with efficiently by moving the boundary to $x=\epsilon$, the $\epsilon$-dependence then canceling at the end
provided the integral is convergent.

This algorithm for computing derivatives can be used to efficiently (and automatically) compute the symbol of any one-dimensional integral;
the result depends only on the symbol of $F$, not on its functional representative.

\section{Special functions for MHV and NMHV hexagons}
\label{app:functions}

The six-dimensional massless hexagon integral
is~\cite{arXiv:1104.2781,arXiv:1104.2787}: \ba
 I_6^{6D}&=&  - \frac13 J^3 - \frac{\pi^2}{3} J +2\sum_{i=1}^3 L^-(x^+u_i, x^-u_i),\nl
 J&=&\sum_{i=1}^3 (\ell_1(x^+u_i)-\ell_1(x^-u_i)),\nl
  x^+ &=& \frac{\l1245\r\l3461\r\l2356\r}{\l 1234\r\l 3456\r\l5612\r}, \quad
  x^- = \frac{\l1245\r\l3461\r\l2356\r}{\l 6123\r\l 4561\r\l2345\r}.
\ea This expression is valid for all Euclidean kinematics,
$u_1,u_2,u_3$ real and positive.
There appear the special combinations introduced in \cite{Goncharov:2010jf} \ba
 L^+_n(x^+,x^-) &=& \frac{\log(x^+x^-)^n}{(2n)!!} + \sum_{m=0}^{n{-}1}  \frac{(-1)^m}{(2m)!!} \log(x^+x^-)^m\left( \ell_{n-m}(x^+) + \ell_{n{-}m}(x^-) \right),
 \nl
  L^-_n(x^+,x^-) &=& \sum_{m=0}^{n{-}1}  \frac{(-1)^m}{(2m)!!} \log(x^+x^-)^m\left( \ell_{n-m}(x^+) - \ell_{n{-}m}(x^-) \right),
\ea
where
\be
 \ell_n(x) = \frac12(\Li_n(x) - (-1)^n \Li_n(1/x)).
\ee
Regarding these special functions,
it is important to note that in our applications the arguments $x^+$ and $x^-$ are the two roots of a quadratic equation with real coefficients.
When the two roots are complex, the standard branch of the polylogarithms is to be used (the one defined on the complex plane minus the line $[1,\infty)$).
When the two roots become real, one is instructed to add opposite infinitesimal quantities $x^+\to x^+ + i\epsilon$, $x^-\to x^- - i\epsilon$; the
hexagon function will not depend on the sign of $\epsilon$.

The collinear limit of the N${}^2$MHV one-loop amplitude gives rise to following seven integrals related
to the two-loop NMHV 6-point amplitudes:
\ba
 \bar Q R_6^\textrm{NMHV} &=&
 (6) \bar Q\log\frac{\l\bar62\r}{\l\bar64\r} f_1
 + ((1)-(2)+(4)-(5))\bar Q\log\frac{\l\bar64\r}{\l\bar62\r}f_2
 + ((2)-(4))\bar Q\log\frac{\l\bar64\r}{\l\bar62\r}f_3\nl &&
 + \left((6) \bar Q\log\frac{\l\bar62\r}{\l\bar63\r}+((5)-(4))\bar Q\log\frac{\l\bar62\r}{\l\bar64\r}\right)f_4
 + ((2)+(4))\bar Q\log\frac{\l\bar62\r}{\l\bar64\r} f_5\nl&&
 + (5) \bar Q\log\frac{\l\bar62\r}{\l\bar63\r}f_6
 + (3) \bar Q\log\frac{\l\bar62\r}{\l\bar63\r}f_7,
\ea
where $(1)$ notates the R-invariant $[23456]$ and
\ba
f_1&=&\frac12 I_{6}^{6D}+\Li_3(1-u_2)+\Li_3(1-1/u_2)-\Li_3(1-u_1)-\Li_3(1-1/u_1)+\frac{\pi^2}{3} \log u_3
\nl&&+\frac{1}{2} (\Li_2(1-u_1)+\Li_2(1-u_2)+\Li_2(1-u_3)) \log \frac{u_1}{u_2 u_3}+\frac{\log^3u_1}{6}-\frac{\log^3u_2}{6}
\nl &&
-\left(\Li_2(1-u_2) +\frac{1}{2} \Li_2(1-u_3)\right) \log u_3
-\frac{3}{4} \log u_1 \log ^2u_3-\frac{1}{4} \log u_2 \log ^2u_3,
\nl f_2&=&
\Li_3(1-u_3)-\frac{1}{2} \left(\Li_2(1-u_3)+\frac{1}{2} \log u_3 \log u_1 u_2-\log u_1 \log u_2\right)\log u_3,
\nl f_3&=&
-\Li_3(1-1/u_3)+\frac16 \log^3u_3+\frac{\pi ^2}{6}  \log u_3,
\nl f_4&=&
\left(\Li_2(1-u_1)+\Li_2(1-u_2)+\Li_2(1-u_3)+\log u_1\log u_3
-\frac{\pi ^2}{3}\right)\log u_3,
\nl f_5&=&
\left(\Li_2(1-u_3)-\frac{\pi ^2}{6}\right) \log \frac{u_1}{u_2},
\quad f_6=
\log ^2u_3 \log \frac{u_2}{u_1},
\quad f_7=
\log u_2 \log u_3 \log \frac{u_3}{u_1}.
\ea
Expressing the six-gluon NMHV amplitude as
\ba
 R_{6,1} &=& \frac12\left( [(1)+(4)]V_3 + [(2)+(5)]V_1 + [(3)+(6)]V_2 \right.\nl
  && \hspace{0.5cm} \left.+  [(1)-(4)]\tilde V_3 + [(5)-(2)]\tilde V_1 + [(3)-(6)]\tilde V_2\right),
\ea
we obtain from this, as explained in the main text, the differential
\ba
 dV_3 &=& -\frac{1}{2} I_6^{6D} d\log\frac{y_2}{y_3}  + (dV_3)_1 d\log\frac{u_1}{(1-u_2)(1-u_3)}
 + (dV_3)_2 d\log\frac{u_2}{1-u_2} + (dV_3)_3 d\log\frac{u_3}{1-u_3}  \nl && +(dV_3)_4  d\log\frac{1-u_1}{u_2 u_3} \nonumber
\ea where \be
y_1=\frac{\l1235\r\l2346\r\l1456\r}{\l1234\r\l2456\r\l1356\r},\quad
y_2=\frac{\l2345\r\l1356\r\l1246\r}{\l1345\r\l2346\r\l1256\r},\quad
y_3=\frac{\l1345\r\l2456\r\l1236\r}{\l1235\r\l3456\r\l1246\r},\nonumber \ee
\ba
 (dV_3)_1 &=&
2 \Li_3(1-u_2)+\Li_3(1-1/u_2)+2 \Li_3(1-u_3)+\Li_3(1-1/u_3)-\frac{1}{6} \log ^3(u_2 u_3)
\nl &&+\Li_2(1-u_1) \log u_2 u_3+\Li_2(1-u_2) \log \frac{u_3^2}{u_1}+\Li_2(1-u_3) \log\frac{u_2^2}{u_1}
\nl&&+\frac{1}{2} \log u_1 \left(\log ^2u_2 u_3+2 \log u_2 \log u_3\right)+\frac{\pi^2}{3} \log \frac{u_1}{u_2^2 u_3^2},
\nl (dV_3)_2 &=&
\Li_3(1-u_1)+\Li_3(1-1/u_1)+\Li_3(1-u_2)+2 \Li_3(1-1/u_2)-3 \Li_3(1-u_3)-2 \Li_3(1-1/u_3)
\nl &&
+\frac12\Li_2(1-u_1) \log \frac{u_2}{u_1u_3^3}+\frac12\Li_2(1-u_2) \log \frac{u_1}{u_2u_3^3}+\frac12\Li_2(1-u_3) \log \frac{u_1^3u_3}{u_2^3} +\frac{\pi ^2}{3}\log u_2 u_3
\nl&&
+\frac13 \log ^3 u_1u_3 -\frac13 \log^3 u_2-\frac12\log u_1u_2 \log^2 u_1u_3+\frac12\log u_1\log u_3\log\frac{u_2^3}{u_3^2}
-\frac{\pi^2}{2} \log \frac{u_1 u_2}{u_3},
\nl (dV_3)_3 &=& (dV)_2 \mbox{ with } u_2\leftrightarrow u_3,
\nl (dV_3)_4&=&
-\left(\Li_2(1-u_1)+\Li_2(1-u_2)+\Li_2(1-u_3)+\log u_1 \log u_2 u_3-\log u_2 \log u_3-\frac{\pi ^2}{3}\right)\log u_1
\nonumber
\ea
and
\ba
 d\tilde V_3&=& \frac12 I_6^{6D} d\log \frac{u_2(1-u_3)}{(1-u_2)u_3} +(d\tilde V_3)_1 d\log y_1 + (d\tilde V_3)_2 d\log y_2 y_3 + (d\tilde V_3)_3 d\log \frac{y_2}{y_3},
\nl
(d\tilde V_3)_1 &=&
\Li_3(1-1/u_2)-\Li_3(1-1/u_3)+\left(\Li_2(1-u_1) -\frac{\pi^2}{3}\right)\log \frac{u_2}{u_3}-\frac{1}{6} \log^3\frac{u_2}{u_3}
\nl
(d\tilde V_3)_2 &=& -\Li_3(1-u_2)+\Li_3(1-u_3)
 +\frac{1}{2} \Li_2(1-u_2) \log\frac{u_1 u_2}{u_3}  -\frac{1}{2} \Li_2(1-u_3) \log \frac{u_1 u_3}{u_2}\nl
 &&
+ \frac{1}{2} \left(\Li_2(1-u_1) +\frac{1}{2} \log u_1 \log u_2 u_3-\frac{\pi^2}{3}\right)\log \frac{u_2}{u_3},
\nl
(d\tilde V_3)_3&=&
\Li_3(1-u_1)+\Li_3(1-1/u_1)- \left(\frac12\Li_2(1-u_1) +\frac{1}{4} \log^2\frac{u_2}{u_3}+\frac{1}{6} \log ^2u_1+\frac{\pi ^2}{6}\right)  \log u_1.
\nonumber
\ea
These differentials are integrable, e.g., $d^2=0$.  The other ones are obtained by cyclic symmetry.

\end{appendix}

\end{document}